\documentclass[twocolumn,preprint]{aastex63}
\usepackage{xspace}
\newcommand{\m}{M87\xspace}

\received{...}
\revised{...}
\accepted{...}

\shorttitle{RadioAstron observations of M87 at 22\,GHz}
\shortauthors{Kim, J.\,-Y. et al.}
\graphicspath{{./}{figures/}}

\begin{document}

\title{\textit{RadioAstron} Space VLBI Imaging of the jet in \m: I. Detection of high brightness temperature at 22\,GHz}

\correspondingauthor{Jae-Young Kim}
\email{jykim@knu.ac.kr}

\author[0000-0001-8229-7183]{Jae-Young Kim}
\affiliation{Department of Astronomy and Atmospheric Sciences, Kyungpook National University, Daegu 702-701, Republic of Korea}
\affiliation{Max-Planck-Institut f\"ur Radioastronomie, Auf dem H\"ugel 69, D-53121 Bonn, Germany}

\author[0000-0001-6214-1085]{Tuomas Savolainen}
\affiliation{Aalto University Department of Electronics and Nanoengineering, PL 15500, FI-00076 Aalto, Finland}
\affiliation{Aalto University Mets\"ahovi Radio Observatory,
             Mets\"ahovintie 114, FI-02540 Kylm\"al\"a, Finland}
\affiliation{Max-Planck-Institut f\"ur Radioastronomie, Auf dem H\"ugel 69, D-53121 Bonn, Germany}

\author[0000-0002-1290-1629]{Petr Voitsik}
\affiliation{Lebedev Physical Institute of the Russian Academy of Sciences, Leninsky prospekt 53, 119991 Moscow, Russia}

\author{Evgeniya V. Kravchenko}
\affiliation{Moscow Institute of Physics and Technology, Dolgoprudny, Institutsky per., 9, Moscow 141700, Russia}
\affiliation{Lebedev Physical Institute of the Russian Academy of Sciences, Leninsky
prospekt 53, 119991 Moscow, Russia}

\author[0000-0001-6088-3819]{Mikhail M. Lisakov}
\affiliation{Max-Planck-Institut f\"ur Radioastronomie, Auf dem H\"ugel 69, D-53121 Bonn, Germany}
\affiliation{Lebedev Physical Institute of the Russian Academy of Sciences, Leninsky prospekt 53, 119991 Moscow, Russia}

\author[0000-0001-9303-3263]{Yuri Y. Kovalev}
\affiliation{Max-Planck-Institut f\"ur Radioastronomie, Auf dem H\"ugel 69, D-53121 Bonn, Germany}
\affiliation{Lebedev Physical Institute of the Russian Academy of Sciences, Leninsky prospekt 53, 119991 Moscow, Russia}
\affiliation{Moscow Institute of Physics and Technology, Dolgoprudny, Institutsky per., 9, Moscow 141700, Russia}

\author{Hendrik M\"uller}
\affiliation{Max-Planck-Institut f\"ur Radioastronomie, Auf dem H\"ugel 69, D-53121 Bonn, Germany}

\author[0000-0003-1622-1484]{Andrei P. Lobanov}
\affiliation{Max-Planck-Institut f\"ur Radioastronomie, Auf dem H\"ugel 69, D-53121 Bonn, Germany}
\affiliation{Moscow Institute of Physics and Technology, Dolgoprudny, Institutsky per., 9, Moscow 141700, Russia}

\author[0000-0001-5991-6863]{Kirill V. Sokolovsky}
\affiliation{ Department of Astronomy, University of Illinois at Urbana-Champaign, 1002 W. Green Street, Urbana, IL 61801, USA}
\affiliation{Center for Data Intensive and Time Domain Astronomy, Department of Physics and Astronomy, Michigan State University, 567 Wilson Rd, East Lansing, MI 48824, USA}
\affiliation{Sternberg Astronomical Institute, Moscow State University, Universitetskij pr. 13, 119992 Moscow, Russia}

\author[0000-0002-5182-6289]{Gabriele Bruni}
\affiliation{INAF-Istituto di Astrofisica e Planetologia Spaziali, via Fosso del Cavaliere 100, I-00133 Roma, Italy}

\author[0000-0002-8186-4753]{Philip G. Edwards}
\affiliation{CSIRO Astronomy and Space Science, PO Box 76, Epping, NSW, 1710, Australia}

\author{Cormac Reynolds}
\affiliation{CSIRO Astronomy and Space Science, PO Box 1130, Bentley, WA 6102, Australia}

\author[0000-0002-7722-8412]{Uwe Bach}
\affiliation{Max-Planck-Institut f\"ur Radioastronomie, Auf dem H\"ugel 69, D-53121 Bonn, Germany}

\author[0000-0002-0694-2459]{Leonid I. Gurvits}
\affiliation{Joint Institute for VLBI ERIC (JIVE), Oude Hoogeveensedijk 4, NL-7991 PD Dwingeloo, the Netherlands}
\affiliation{Faculty of Aerospace Engineering, Delft University of Technology, Kluyverweg 1, NL-2629 HS Delft, the Netherlands}

\author[0000-0002-4892-9586]{Thomas P. Krichbaum}
\affiliation{Max-Planck-Institut f\"ur Radioastronomie, Auf dem H\"ugel 69, D-53121 Bonn, Germany}

\author[0000-0001-6906-772X]{Kazuhiro Hada}
\affiliation{Mizusawa VLBI Observatory, National Astronomical Observatory of Japan, 2-12 Hoshigaoka, Mizusawa, Oshu, Iwate 023-0861, Japan}
\affiliation{Department of Astronomical Science, The Graduate University for Advanced Studies (SOKENDAI), 2-21-1 Osawa, Mitaka, Tokyo 181-8588, Japan}

\author[0000-0002-8657-8852]{Marcello Giroletti}
\affiliation{Istituto Nazionale di Astrofisica, Istituto di Radioastronomia (IRA), via Gobetti 101, 40129, Bologna, Italy}

\author{Monica Orienti}
\affiliation{Istituto Nazionale di Astrofisica, Istituto di Radioastronomia (IRA), via Gobetti 101, 40129, Bologna, Italy}

\author[0000-0002-5989-8498]{James M. Anderson}
\affiliation{Leibniz Institute for Agricultural Engineering and Bioeconomy, 
Max-Eyth-Allee 100,
D-14469 Potsdam-Bornim,
Germany
}

\author{Sang-Sung Lee}
\affiliation{Korea Astronomy and Space Science Institute, 776 Daedeok-daero, Yuseong-gu, Daejeon 34055, Republic of Korea}

\author[0000-0002-4148-8378]{Bong Won Sohn}
\affiliation{Korea Astronomy and Space Science Institute, 776 Daedeok-daero, Yuseong-gu, Daejeon 34055, Republic of Korea}

\author[0000-0001-7470-3321]{J. Anton Zensus}
\affiliation{Max-Planck-Institut f\"ur Radioastronomie, Auf dem H\"ugel 69, D-53121 Bonn, Germany}






\begin{abstract}

We present results from the first 22\,GHz space very-long-baseline interferometric (VLBI) imaging observations of \m by {\it RadioAstron}. As a part of the Nearby AGN Key Science Program, the source was observed in Feb 2014 at 22\,GHz with 21 ground stations, reaching projected $(u,v)$-spacings up to $\sim11\,$G$\lambda$. 
The imaging experiment was complemented by snapshot {\it RadioAstron} data of \m obtained during 2013--2016 from the AGN Survey Key Science Program. Their longest baselines extend up to $\sim25\,$G$\lambda$. 
For all these measurements, fringes are detected only up to $\sim$2.8\,Earth Diameter or $\sim$3\,G$\lambda$ baseline lengths, resulting in a new image with angular resolution of $\sim150\,\mu$as or $\sim20$\,Schwarzschild radii spatial resolution.
The new image not only shows edge-brightened jet and counterjet structures down to submilliarcsecond scales but also clearly resolves the VLBI core region. While the overall size of the core is comparable to those reported in the literature, the ground-space fringe detection and slightly super-resolved {\it RadioAstron} image suggest the presence of substructures in the nucleus, whose minimum brightness temperature exceeds $T_{\rm B, min}\sim10^{12}\,$K. 
It is challenging to explain the origin of this record-high $T_{\rm B, min}$ value for \m by pure Doppler boosting effect with a simple conical jet geometry and known jet speed.
Therefore, this can be evidence for more extreme Doppler boosting due to a blazar-like small jet viewing angle or highly efficient particle acceleration processes occurring already at the base of the outflow.

\end{abstract}

\keywords{
Galaxies: jets -- 
Galaxies: active -- 
Galaxies: individual: M87 -- 
Techniques: interferometric -- 
Techniques: high angular resolution
}


\section{Introduction}

A certain fraction of accreting supermassive black holes (SMBHs) in Active Galactic Nuclei (AGN) launch powerful and collimated beams of plasma, which are referred to as jets
(see, e.g., \citealt{blandford19}). Theoretical studies suggest that those AGN jets can be launched by extraction of the energy of the spinning black hole or the inner accretion disk \citep{bz77,bp82}. While these models are found to be promising in numerical simulations of black hole accretion systems, especially with strong magnetic field strengths (see, e.g., \citealt{yuan14} for a review), observational tests of the jet formation models have been restricted to only a handful of sources, due to the extremely compact sizes of the vicinity of the black hole.
  
\m (Virgo A, 1228+126, NGC 4486, 3C\,274B) is a nearby giant radio galaxy, located at a  luminosity distance of only $d_{\rm L}=16.8\,$Mpc \citep{eht2019, blakeslee09, bird2010} with a central supermassive black hole of mass $M_{\odot}=6.5\times10^{9}M_{\rm sun}$ \citep{eht2019}.
This combination gives an angular-to-spatial resolution conversion factor of 0.08\,pc or 131 Schwarzschild radii ($R_{\rm s}$) per 1 milliarcsecond (mas), or 2.61$R_{\rm s}$ per $20$ microarcseconds ($\mu$as), providing the best opportunity to probe
the compact jet launching region down to the event horizon scales (see, e.g., \citealt{blandford19}). 
In this regard, the very-long-baseline interferometry (VLBI) technique uniquely offers imaging capability at the highest angular resolution and therefore has been a crucial tool to directly image the mass accretion and jet launching in \m.

Since the early detection and imaging of the compact core and radio jet in \m by VLBI technique (e.g., \citealt{reid82}), the source has been observed by modern VLBI facilities, to study the structure, dynamics, and physical origin of the jet (see, e.g., \citealt{eht2019} and references therein).
Along with those studies, decades-long efforts have been made to improve angular resolution towards \m by adopting orbiting antenna as a VLBI station, thus realizing a space VLBI array with a virtual aperture larger than the Earth diameter (see, e.g., \citealt{burke09,schilizzi12,hirabayashi12,gurvits20,gurvits21} for a review).
Previous space VLBI programs including 
tests with the NASA Tracking and Data Relay Satellite System (TDRSS; \citealt{levy86,levy89,linfield89,linfield90}) and 
the VLBI Space Observatory Programme (VSOP; \citealt{hirabayashi98,hirabayashi00}) provided not only proof of the concept but also the possibility of imaging \m at exceptionally high angular resolution (e.g., \citealt{dodson06}).
However, these spacecrafts were able to observe and downlink data together only at relatively low Earth orbits (e.g., $<4$ Earth diameters), comparatively low observing frequencies of $\lesssim 15\,$GHz, and the limited sensitivity, thus offering only limited angular resolutions.

{\it RadioAstron} is the latest space VLBI mission dedicated for ultra-high resolution VLBI observations at radio observing frequencies of 1.6 to 22\,GHz, using the space radio telescope \textit{Spektr-R} \citep{kardashev13}. Thanks to the excellent antenna tracking and orbit determination capability, interferometric fringes of {\it RadioAstron} have been detected
on baseline lengths up to $\sim 28$ Earth diameters \citep{kovalev20}.
As part of the mission, a key science program on nearby radio galaxies has been focusing on producing the sharpest images of nearest accreting black holes up to angular resolutions of $\sim30\mu$as (see \citealt{giovannini18} and \citealt{savolainen21}).
Among various targets in the key science program, \m was observed by {\it RadioAstron} in Feb 2014, 
in full-track global VLBI and up to $\sim11$G$\lambda$ baseline length at  22\,GHz,
in order to resolve the complex structure of the jet and its origin, down to the event-horizon-scale at $\sim20\,\mu$as angular resolution.
In this paper, we present the first results from this {\it RadioAstron} observation of \m at 22\,GHz.
Also, \m was more frequently observed by {\it RadioAstron} through the AGN brightness temperature survey program \citep{kovalev20} in snapshot mode, which spans 
the years 2013--2016 and baseline lengths up to $\sim25\,$G$\lambda$. Results from this program are also presented.

The paper is organized as follows.
In \S\ref{sec:whole_datareduction} and \S\ref{sec:analysis} we describe details of the reduction and analysis of data from {\it RadioAstron} and other accompanying observations. The main results, including the highest-angular resolution image of \m at 22\,GHz, are shown in \S\ref{sec:results}. 
We discuss the major implications of the findings in \S\ref{sec:discussions} and conclude our study in  \S\ref{sec:conclusions}.

\section{Observations, Data Reduction, and Imaging}\label{sec:whole_datareduction}

\subsection{RadioAstron 22\,GHz}\label{sec:ra_datareduction}

\begin{figure}
    \includegraphics[width=0.45\textwidth]{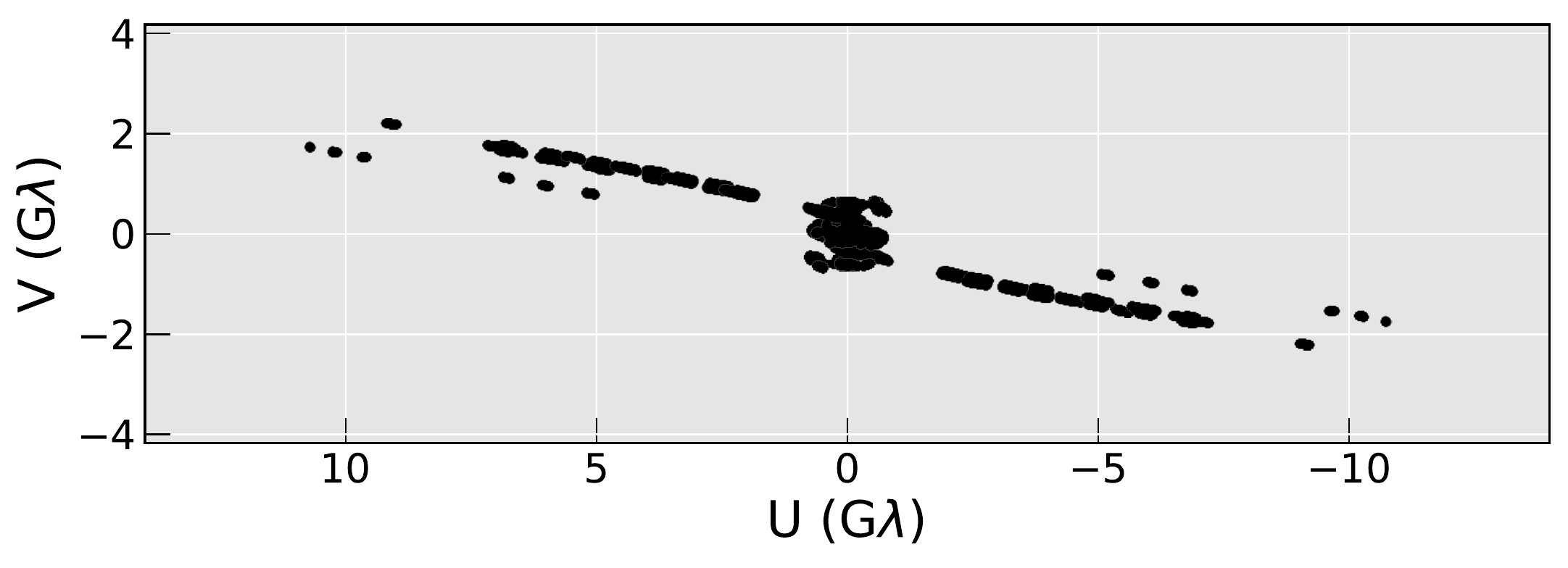}
    \caption{
    $(u,v)$-coverage of the scheduled {\it RadioAstron} observations of \m at 22\,GHz (i.e., before the fringe detection).}
    \label{fig:uvcoverage_raw}
\end{figure}

\begin{figure*}[ht!]
    \gridline{
    \fig{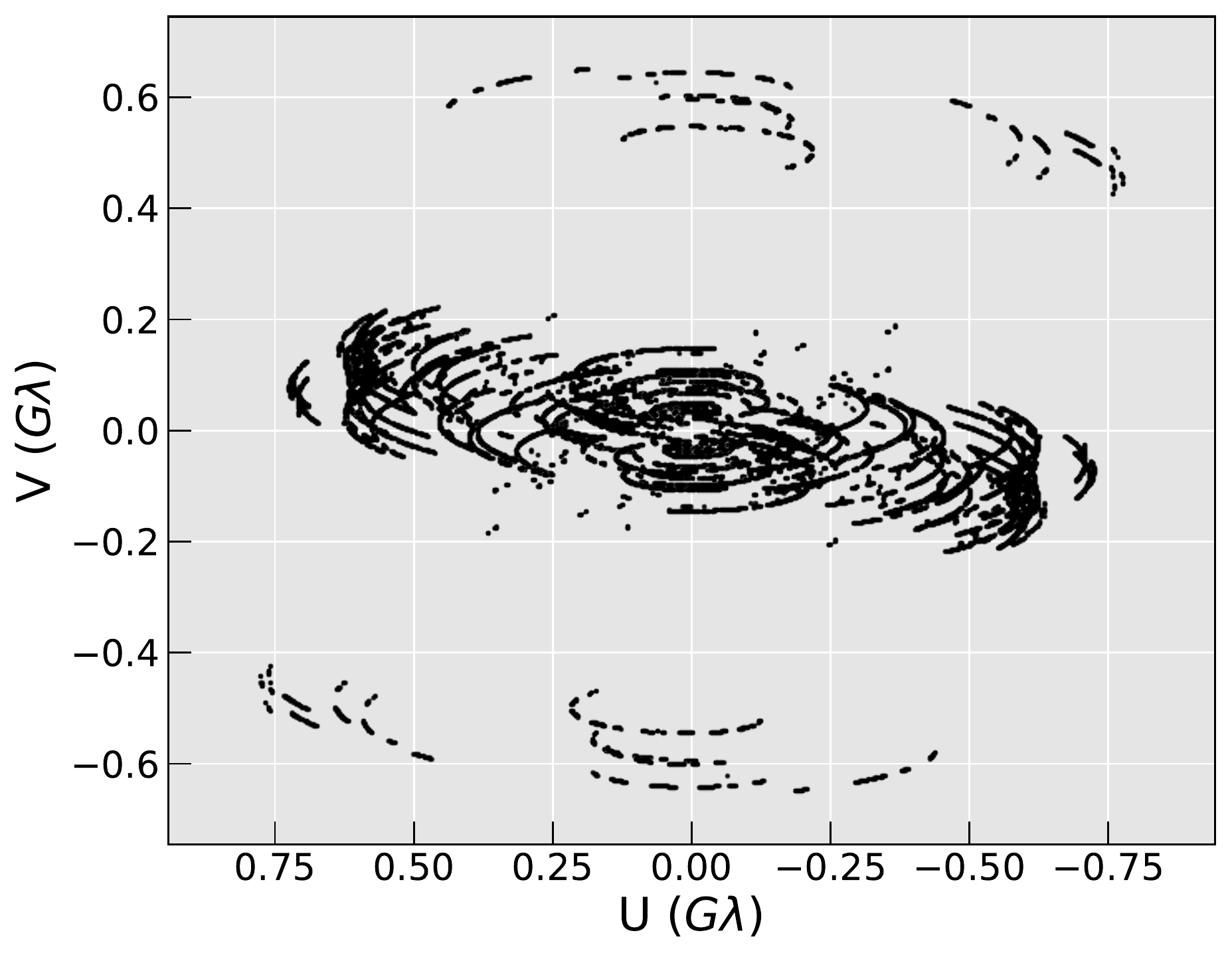}{0.5\textwidth}{}
    \fig{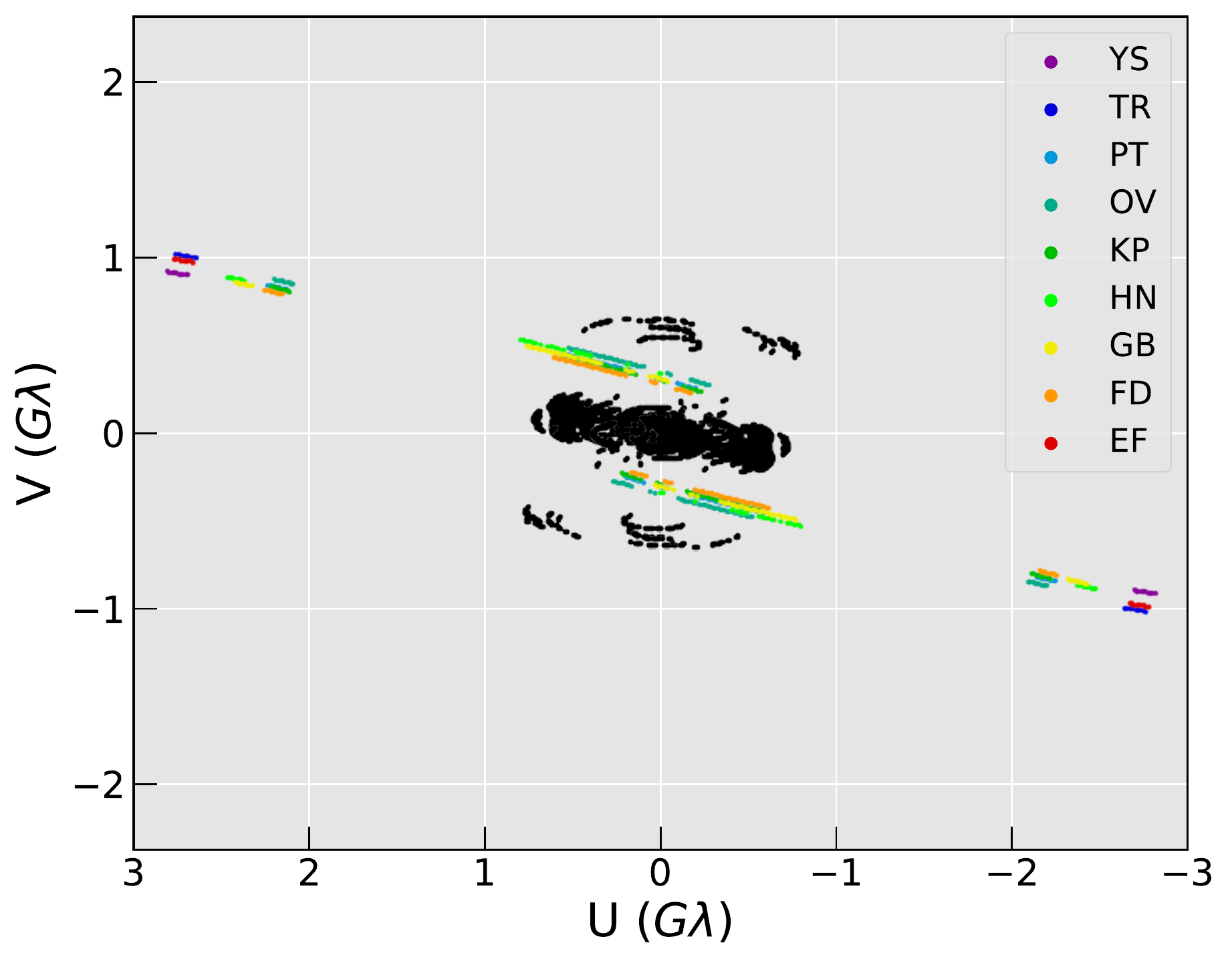}{0.5\textwidth}{}
    }
    \caption{
    $(u,v)$-coverage of the {\it RadioAstron} observations of \m at 22\,GHz on 2014 Feb 4--5, after post-correlation data processing and imaging with self-calibration. 
    {\it Left:} Ground-only $(u,v)$-coverage.
    {\it Right:} Full coverage (ground+space) with space baselines whose fringes are detected. Ground-to-space baselines are color-coded for each ground station (see Table \ref{tab:station_list} for the station code).
    }
    \label{fig:uvcoverage}
\end{figure*}

\m was observed by the {\it RadioAstron} mission at 22.236\,GHz ($\lambda=1.3$\,cm; K-band) from 2014 Feb 4, 16:00 to Feb 5, 12:53, 2014, UT, 
as a part of the Nearby AGN Key Science Program (experiment code raks03b; gs032; see, e.g., \citealt{bruni20} for the description of the program).
The ground-array consisted of in total 21 telescopes. Their names, VLBI station code, antenna diameters, and SEFDs estimated from the station DPFU and system temperatures during the imaging observations are listed in Table\,\ref{tab:station_list}.
The ground stations observed the source and calibrators 1226+032 (3C\,273) and PKS\,1236+077 
in both left- and right-handed circular polarizations (LCP and RCP, respectively), at an central observing frequency of 22.236\,GHz with a total bandwidth of 32\,MHz (total data bit rate of 256\,Mbps with 2-bit sampling) using two intermediate frequency (IF) bands (thus 16\,MHz per IF per polarization). 
The Space Radio Telescope (SRT) simultaneously observed 
 \m at both 5\,GHz and 22\,GHz bands, with the same 64\,MHz total bandwidth (total data bit rate of 128\,Mbps with 1-bit sampling), using two IFs at each band and only in LCP. This setup resulted in the data bandwidth of 16\,MHz per IF per polarization. We refer to Kravchenko et al. (in preparation) for the reduction, analysis, and discussions of the 5\,GHz band data.
The maximum distance to the SRT was $\sim$11.5 Earth diameter ($D_{\rm Earth}$), corresponding to the fringe spacing of $\sim10.9$G$\lambda$ at our observing frequency. 
The projected $(u,v)$-coverage is shown in Figure~\ref{fig:uvcoverage_raw}.  
We note the highly elongated orbit of the spacecraft in the E-W direction, which gives higher resolution along the direction of the nearly E-W oriented jet in \m87.

\begin{figure}[ht!]
    \includegraphics[width=0.45\textwidth]{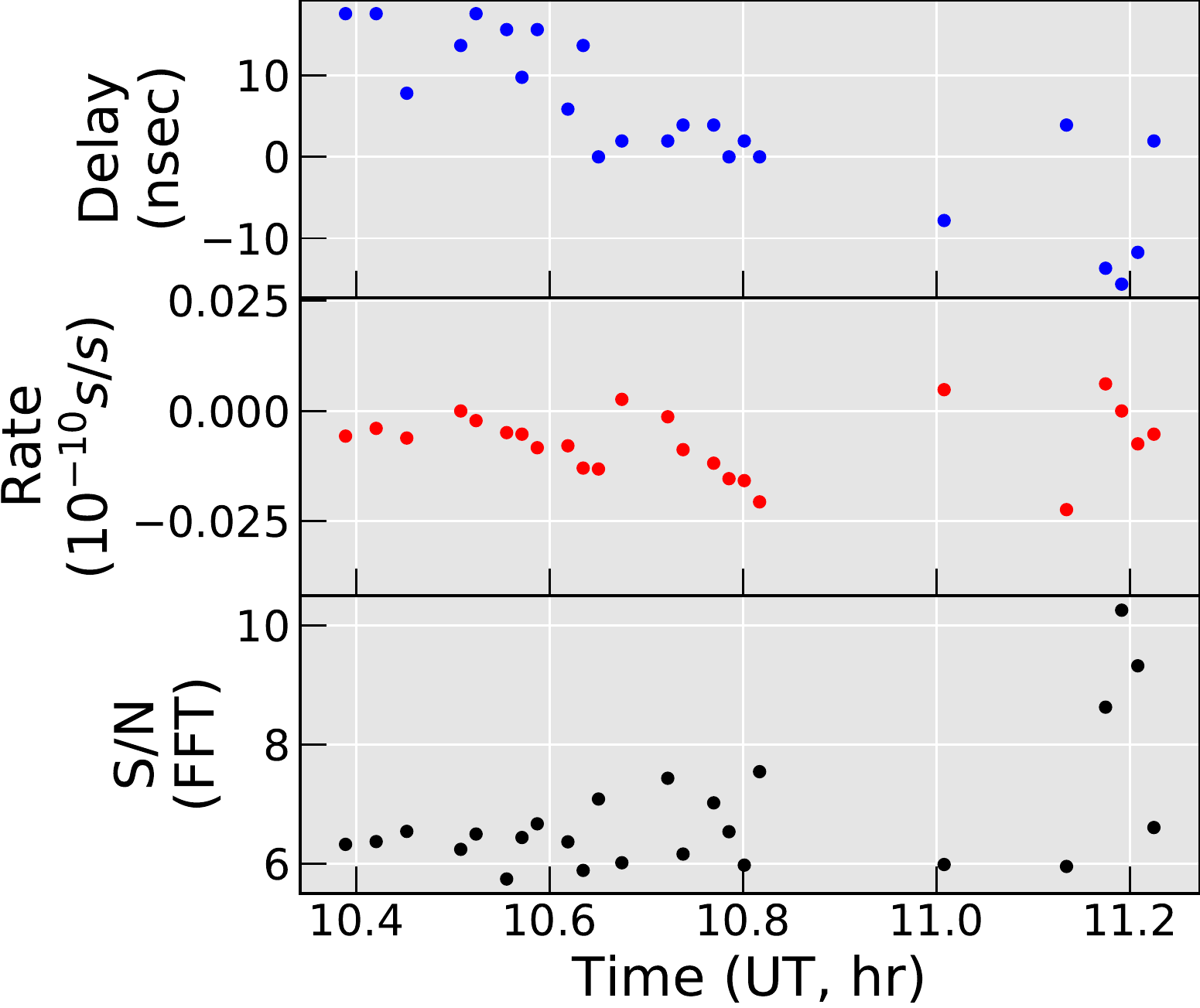}
    \caption{
    Fringe solutions for short ($\lesssim1\,D_{\rm Earth}$) space baselines from AIPS FRING with solution interval of 1~min. We note time evolution of the residual fringe delays and rates, which most likely indicates the presence of acceleration of the SRT not solved for in the data correlation.
    }
    \label{fig:fringe}
\end{figure}

\begin{table}[]
    \centering
    \begin{tabular}{llll}
        \hline
        Name & Station Code & Diameter & SEFD \\
             &      & (m) & (Jy) \\
        \hline
ATCA (5$\times$22\,m)$^{a}$ &  AT    &  49 & 101 \\
Badary                      &  BD    &  32 & 1000 \\
VLBA--Brewster              &  BR    &  25 & 536 \\
Effelsberg                  &  EF    & 100 & 160 \\
VLBA--Fort Davis            &  FD    &  25 & 730 \\
Green Bank                  &  GB    & 100 & 23 \\
Hartebeesthoek              &  HH    &  26 & 4500 \\
VLBA--Hancock               &  HN    &  25 & 1087 \\
VLBA--Kitt Peak             &  KP    &  25 & 667 \\
Kalyazin                    &  KZ    &  64 & --$^{b}$ \\
VLBA--Los Alamos            &  LA    &  25 & 574 \\
VLBA-Mauna Kea              &  MK    &  25 & 843 \\
Mopra                       &  MP    &  22 & 1000 \\
VLBA--North Liberty         &  NL    &  25 & 926 \\
VLBA--Owens Valley          &  OV    &  25 & 926 \\ 
VLBA--Pie Town              &  PT    &  25 & 642 \\
$Spektr-R$ SRT              &  RA    &  10 & 44160 \\
VLBA--Saint Croix           &  SC    &  25 & 550 \\
Svetloe                     &  SV    &  32 & 1250 \\   
Torun                       &  TR    &  32 & 880 \\
Yebes                       &  YS    &  40 & 1152 \\
Zelenchukskaya              &  ZC    &  32 & 877 \\
        \hline
    \end{tabular}
    \caption{List of stations forming the ground array and their properties. The SEFD values were calculated based on values of the station DPFU and characteristic system temperatures from the actual imaging observations.
    (a) For ATCA, in total five individual stations were phased up, forming an effective 49\,m diameter dish.
    (b) Kalyazin couldn't record the data in the imaging experiment and therefore no SEFD value is presented in this table.
    }
    \label{tab:station_list}
\end{table}

The observed raw data were correlated at the Max-Planck-Institut f\"ur Radioastronomie using the DiFX correlator \citep{deller11}, adjusted for space VLBI in order to account for the special and general relativistic effects related to the orbiting antenna \citep{bruni16}.
Before fully processing the post-correlation dataset, we first examined the quality of ground-to-space baseline fringes using PIMA \citep{petrov11}, by making use of the baseline-based fringe algorithm, to inspect residuals of the fringe rates due to the acceleration of the orbiting antenna.
We note that PIMA has unique advantages for calibrating the space baselines.
Specifically,  
the program can accurately determine the space-antenna acceleration term and thus significantly constrain ranges of the residual fringe solutions in the subsequent calibration.
For this reason, PIMA has been routinely used for post-correlation calibration of {\it RadioAstron} observations.
Fringes were clearly detected at $<1D_{\rm Earth}$ baselines at PIMA S/N values of $\sim8-72$, while there was no clear initial fringe detection at longer $(u,v)$ spacings\footnote{Here we particularly note that PIMA calculates the fringe S/N values differently from AIPS, and their S/N values are explicitly distinguished in the following discussions.}. 
After this examination, the post-correlation data were loaded into the Astronomical Image Processing System (AIPS) software \citep{greisen03} in order to perform the full a-priori calibrations and, more importantly, to improve the fringe detection for ground-space baselines, taking the advantage of the antenna-based global fringe fitting algorithm which can stack baselines after phase-calibrating the ground stations (see other literature for further details, e.g., \citealt{tms, gomez16,savolainen21}).
We note that the subsequent AIPS calibration still made use of the PIMA results in the cross-comparison of the fringe solutions for robust fringe detections.
In the following, we describe the details of the AIPS data calibration.
To start with, we only calibrated the ground-based stations in order to obtain a ground-only image of \m that can provide a source model for improved fringe fitting of the space baselines (see, e.g., \citealt{giovannini18,savolainen21}).
In particular, the manual phase and delay offsets of each ground antenna were determined using high S/N scans on the calibrators. 
Already at this early stage, we could not find any fringes to AT and MP. Accordingly, these stations were dropped out from further analysis.
Then, a global fringe fitting was performed, using the AIPS FRING task, setting a S/N threshold of 4.5. Solutions were successfully found for most of the scans. 
The left panel of Figure~\ref{fig:uvcoverage} shows the fringe detection and the corresponding $(u,v)-$coverage of the ground array.
The a-priori amplitude calibration was performed using the AIPS task APCAL, based on the aperture efficiencies and system temperature measurements.
Bandpasses of ground stations were also calibrated using their autocorrelation power spectra.

The ground-array only data were exported outside AIPS for imaging of \m in the Difmap software \citep{shepherd94} using the CLEAN algorithm. 
By making iterative use of the CLEAN task and phase and amplitude self-calibrations, we obtained an image of the jet of \m at an angular resolution of $0.19\times0.30$\,mas ($0.22\times0.70$\,mas) at the beam  major axis position angle of  $-1^{\circ}$ ($-7^{\circ}$) with uniform (natural) weighting.

\begin{figure}[t]
    \plotone{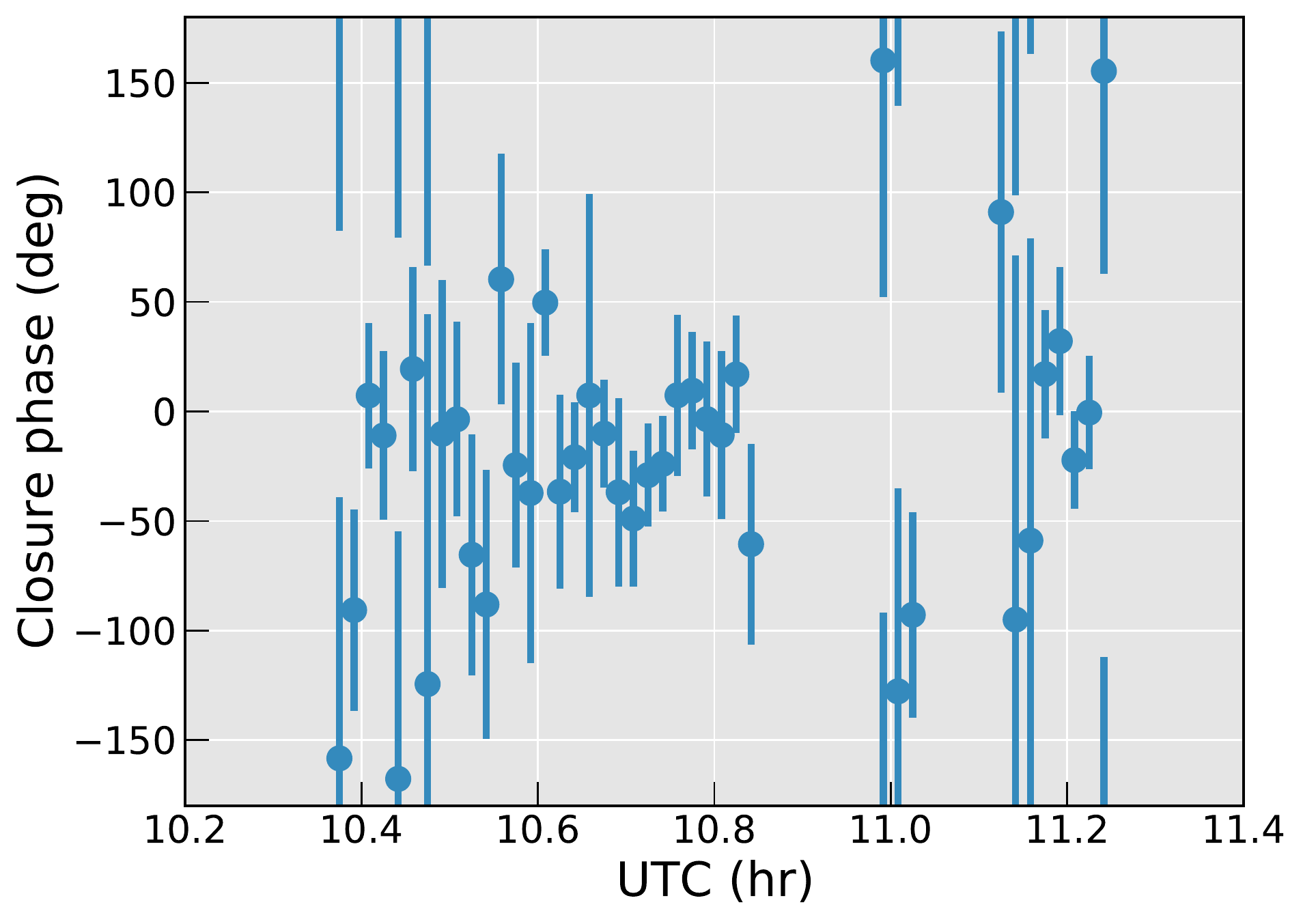}
    \plotone{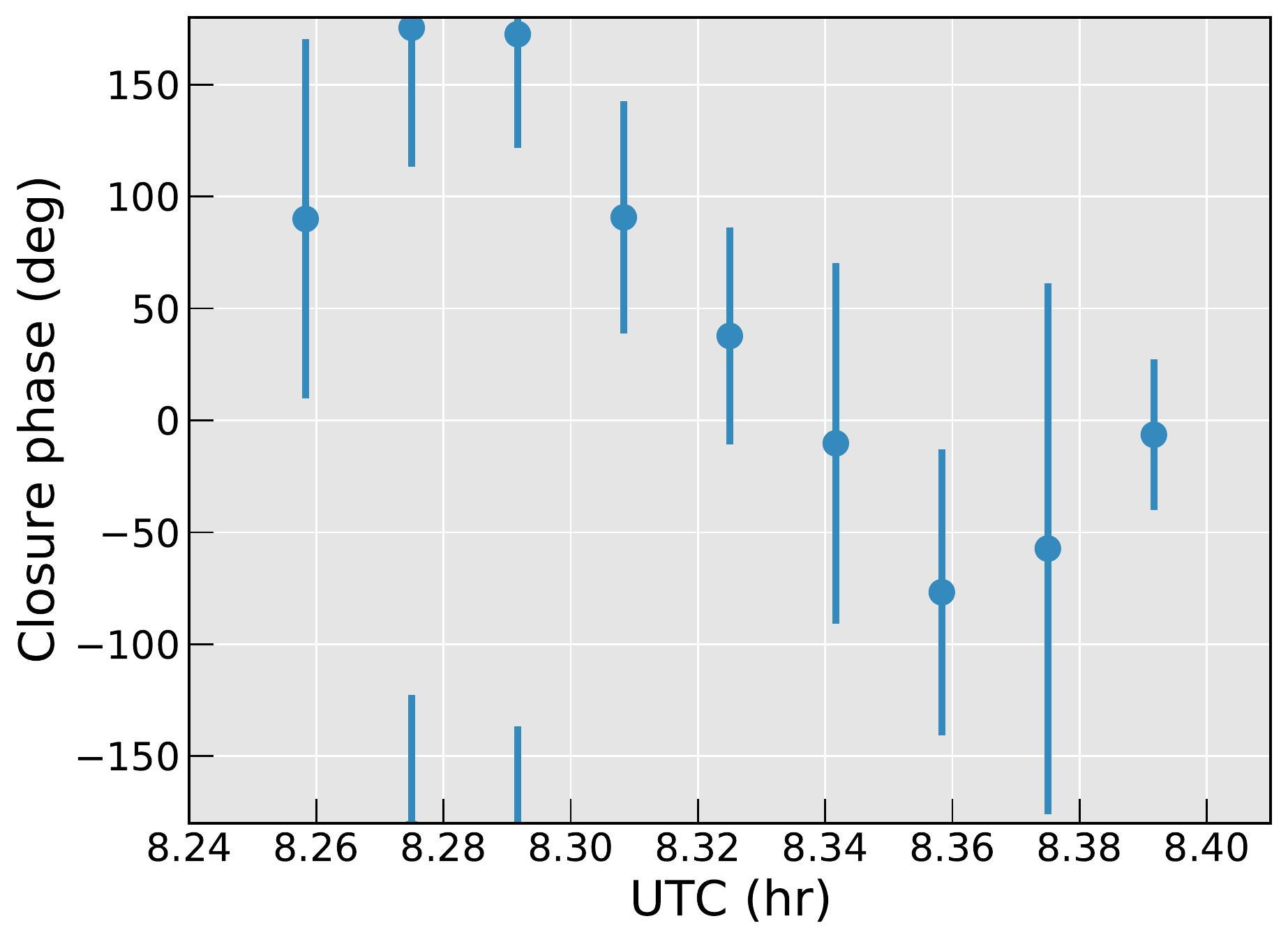}
    \caption{
    Example closure phases of triangles with high S/N and including the SRT.
    {\it Top:} RA-EF-YS triangle for baselines to the SRT of $<1D_{\rm Earth}$.
    {\it Bottom:} RA-EF-GB triangle for baselines to the SRT of $\sim2.9D_{\rm Earth}$.
    In both panels, the data have been averaged over 60\,s without a-priori phase self-calibrations.
    Also note that both panels have different time ranges.
    }
    \label{fig:CP_2.9ED}
\end{figure}

This image was loaded back into AIPS and later used as a model for the global fringe fitting of the SRT with respect to the sensitive reference antennas EF and GB (see also other literature, e.g., \citealt{gomez16,giovannini18,savolainen21}, for the details of calibrating the SRT).
Before fringe-fitting the SRT, we applied additional global fringe fitting of only the ground stations with a short solution interval of 1 minute by AIPS FRING, using the ground-only \m jet image as a model. This step was meant to self-calibrate rapid phase fluctuations due to the atmosphere and thus improve the coherence timescales for baselines to the SRT (i.e., over several minutes).
Next, we performed manual phase-calibrations of the SRT by selecting scans on \m at short $(u,v)$-spacing ($<1D_{\rm ED}$; near the perigee), 
successfully obtaining suitable delay solutions at both IFs. 

After connecting the two IFs in phase, we began searching for fringe solutions scan-by-scan using AIPS FRING by combining the two IFs, starting from short to long baselines, with progressively increasing fringe solution interval of 1 to 10 minutes. This implementation was necessary to take into account the acceleration of the SRT at short $(u,v)$-spacings (e.g., $<1D_{\rm Earth}$), which in turn introduces large phase and delay rate drifts and practically limits the coherence time. 
In addition, we took the advantage of the phased-calibrated ground-array to stack ground baselines connecting the reference stations EF and GB to SRT, setting AIPS FRING dparm(1)=3.
To examine the quality of the fringe solutions and judge on the fringe detection, we particularly searched for smooth and continuous change of the delays and delay rates in time (also in the $(u,v)$-distance) so that we can avoid selecting spurious peaks in the fringe search window.
At the short $<1D_{\rm Earth}$ distances to the SRT, we successfully obtained continuously varying fringe solutions with the AIPS fringe peak S/N of $>6$ and consistently small residual delays and delay rates at both IFs.
Here AIPS S/N refers to the peak-to-noise ratio from the initial baseline-based signal search with simple fast Fourier Transform (FFT), and the final, global fringe S/N after the least square stage is higher (see, e.g., \citealt{schwab83}.
Right panel of Figure~\ref{fig:uvcoverage} shows the space fringe detections at the $<1D_{\rm Earth}$ baselines.
We also show the time evolution of the fringe solutions in Figure~\ref{fig:fringe}.
The time-evolving rate and jumps over the scans indicate clear fringe residuals due to the acceleration of the spacecraft near the perigee.
These fringe delays and rates, obtained from AIPS, were also consistent with values from the initial PIMA fringe search, providing further confidence on the source detection.

At $>1D_{\rm Earth}$, we first calibrated away a large delay offset of $\sim1\,$microsecond for the SRT, which was already known from simultaneous {\it RadioAstron} 4.8\,GHz  observations of \m (Kravchenko et al., in prep.). Then, we adopted the full length of each scan, $\sim10$\,min, as the solution interval of AIPS FRING with a S/N threshold of 3, using narrow delay and rate windows. From this, we were able to find fringes to the SRT up to $\sim2.8D_{\rm Earth}$ baseline length ($\sim$3$G\lambda)$ at an AIPS initial FFT S/N value of 3.4.
The corresponding improvement of these detections for the final ($u,v)$-coverage are highlighted in color in right panel of Figure~\ref{fig:uvcoverage}.
The significance of this detection was rigorously tested in multiple ways, including inspection of how much the fringe solutions in two different IFs were consistent and quantitative analysis of the probability of false fringe detection using a dedicated Monte Carlo simulation. 
The latter was performed following \cite{petrov11} and \cite{savolainen21}.  More details of the Monte Carlo simulation are presented in Appendix~\ref{appendix:pfd}.
After these tests, we were confident with the source detection at this $(u,v)$ spacing, with the false fringe detection probability of $<10^{-4}$. 
We also show in Figure~\ref{fig:CP_2.9ED} example closure phases of the fringe-detected scans.
Upper panel shows closure phases in the small triangle whose values are centered nearly around $\sim0^{\circ}$ with some scatter, indicating a symmetric source structure. The zero closure phases can be due to the bright Gaussian-shaped VLBI core, which is symmetric. On the other hand, clear and non-zero closure phases in the large triangle (bottom panel) suggest the presence of a complex source structure within the beam of the ground-array.

At even longer baseline lengths ($>3D_{\rm Earth})$, we could not find any reliable fringe-fit solutions to the SRT, and thus concluded $>3D_{\rm Earth}$ scans were non-detections.
We also tried to apply bandpass corrections for the SRT, but we did not  obtain reasonable solutions and those solutions did not improve the fringe detection rates in subsequent attempts to fringe-fit the data. Therefore,  we did not correct for the bandpass of the SRT.

The fully calibrated dataset was finally exported from AIPS for imaging with Difmap.
First, we loaded the dataset into Difmap and applied amplitude self-calibration loops to the ground-only array using the CLEAN model of the ground-only image, to prepare a final imaging-ready dataset.
The ground-only model was removed at this point, to take the advantage of the higher angular resolution with the addition of the SRT. We then applied iterative CLEAN and phase-only self-calibrations, with the selfcal solution interval down to 1\,min for the SRT (i.e., equivalent to that of AIPS FRING) and adopting the CLEAN windows of the ground-only model. The latter especially helped avoid over-subtraction of spurious CLEAN components in the counterjet region.
In the course of imaging, we tested various combinations of parameters in Difmap, in particular the $(u,v)$-weighting by choosing UVWEIGHT=$(0,-1), (2,0), (5,0), (2,-1),$ and $(5,-1)$ (the third and last so-called super-uniform  weighting; see \citealt{gomez16,giovannini18,savolainen21}).
We found that the choice of $(2,-1)$ provided the best compromise for the angular resolution, image sensitivity, and overall reliability of the resulting image against over-fitting thermal noises and systematic sidelobes.
We also attempted to self-calibrate the amplitudes of the SRT but only deriving a single constant correction factor, in order to investigate the accuracy of the amplitude calibration of the SRT. The self-calibration resulted in $\lesssim$10\% change in the amplitude of the SRT. This correction neither changed the image quality nor major features in the image significantly. Therefore, we did not apply further amplitude self-calibration solutions.

Before determining the final image, we examined how reliable detailed features in the map by (i) imaging the real dataset by different authors without their mutual interactions about the images and (ii) creating and imaging a realistic synthetic dataset, which adopted the same S/N ratio and $(u,v)$-coverage as the real observation, and also adopted a known ground-truth source image. 
Details of the synthetic data generation and imaging tests are described in Appendix \ref{appendix:synthetic_data}.
These tests provided confidence that the presence of the counterjet, well resolved nucleus of the jet, and their overall structure are reliable.
In the end, a final image was produced at 
an angular resolution of $0.15\times0.47$\,mas with \texttt{uvweight$=(2,-1)$} at a beam position angle (PA) of $-16^{\circ}$.
For completeness information, we also report the beam sizes corresponding to the uniform and super-uniform weightings from 
\texttt{uvweight=(2,0)}, 
\texttt{(5,0)}, and 
\texttt{(5,-1)}, which are 
$0.10\times0.32$\,mas at PA $-20.2^{\circ}$, 
$0.09\times0.28$\,mas at PA $-20.4^{\circ}$, and  
$0.14\times0.41$\,mas at PA $-16.3^{\circ}$, respectively.

\subsection{VLBA 43\,GHz}

For comprehensive multiwavelength analysis of the jet in \m, the source was also observed by the Very Long Baseline Array (VLBA) on Feb 05, 2014 at 08:09$-$12:46 UT, that was interleaved with {\it RadioAstron} scans (see \S\ref{sec:ra_datareduction}). The VLBA observation was configured with the central observing frequency of 43.136\,GHz with a total bandwidth of 64\,MHz in two circular polarizations (LCP and RCP), split between two 16\,MHz IFs.
Standard procedures were similarly applied to fringe-fit the data and calibrate the amplitudes in AIPS, and to image the source structure in Difmap. For the imaging in Stokes $I$, the data were averaged over 30\,s in time; see descriptions for the ground-array data in \S\ref{sec:ra_datareduction}. 
A final image was obtained at an angular resolution of 0.15$\times$0.37\,mas with the beam position angle of $-1.6^{\circ}$.

\subsection{{\it RadioAstron} 22\,GHz AGN survey}\label{subsec:ra_survey_data}

Besides the imaging experiment, \m was regularly monitored by {\it RadioAstron} as part of the AGN survey program (see \citealt{kovalev20}).
The survey observations were performed in the visibility tracking (non-imaging) mode with few ground stations participating in each experiment. If fringes to the space baselines are detected, comparing the correlated flux densities measured at ground-ground and ground-space baselines to a simple Gaussian source model can provide estimates for the angular size and brightness temperature of a source from such snapshot observation.
The survey observations expand to much longer in the $(u,v)$ space, up to $\sim25\,$G$\lambda$.
Those observations were made during 2013--2016 and therefore it is not straightforward to directly combine those complex visibilities with the 22\,GHz imaging data. In spite of the limitation, however, the survey data still provide significant constraints on the presence or absence of any small-scale structure in the jet of \m in the time-averaged sense.
Also, adding the survey data to the imaging observation can potentially boost the angular resolution of the final image (e.g., \citealt{gomez22}).
Therefore, we also adopted those datasets to search for the  most compact structure in \m at 22\,GHz.
 More details of those data, including the scheduling, observation, data reduction, and analysis can be found in \cite{kovalev20}.

\section{Analysis}\label{sec:analysis}

\subsection{Modeling the VLBI core}\label{subsec:modelfit}

In this work, we identify the VLBI core as the most compact and brightest component at the base of the approaching jet, which is 
the location of the intensity peak.
To characterize the basic properties of this region, we fitted an elliptical Gaussian to the region close to the peak of the intensity. Before doing so, we subtracted from the visibility CLEAN components outside $\pm0.3$\,mas and $\pm0.1$\,mas from the peak in RA and Dec, to avoid fitting the extended jet emission by the core Gaussian model. We note that an elliptical Gaussian is more reliable than fitting several smaller circular Gaussians, when defining the overall size of the nucleus and estimating brightness temperature, since the latter can be strongly affected by the size estimate \citep[e.g.,][]{kovalev05,jorstad17}. 
For the actual fitting, we primarily use the modelfit procedure implemented in Difmap to fit the Gaussian to the long-baseline visibilities. 
We note that statistical uncertainties of the fitted parameters can be in principle determined using the parameters and residuals of the fit \citep[e.g.,][]{fomalont99}. 
However, the underlying systematic uncertainties can be significantly larger in our case, due to the highly elongated $(u,v)$ coverage of the space VLBI observation. Therefore, we performed an independent fit of the elliptical Gaussian in the image domain,  using the JMFIT task in AIPS  \citep[see, e.g.,][]{hada13,hada16}. 
We note that the image plane method can also well describe the overall nuclear region, as long as the main interest is the overall size and flux density of a feature and not the fine-scale subnuclear structures that are often smaller than the nominal beam. 

After having performed these fitting procedures and obtained the parameters, we took the average of the fitted parameters from the image and visibility and half of their differences as the values and associated uncertainties of the Gaussian modelfit parameters.
This approach resulted in the following uncertainties for the fitted parameters: $\sim10\%$ (0.09\,Jy) for the flux density, $\sim11\%$ (0.04\,mas) and $\sim8\%$ (0.01\,mas) for the sizes along the major and minor axes respectively, and accordingly $\sim18\%$ for the brightness temperature (see \S\ref{sec:results}).
The $\sim10\%$ flux error is slightly larger than the usual flux uncertainties of $\sim5\%$ for the Gaussian modeling of well-separated jet components in typical cm-VLBI observations (e.g., \citealt{lister09}). On the other hand, the size uncertainties are of the order of $\sim10\%$ of the space VLBI beam size, which might still be slightly underestimated compared to $\sim20\%$ beam sizes that were suggested as characteristic size errors by other studies (see, e.g., \citealt{homan02,lister09,jorstad17} and references therein).
We note that adopting the larger 20\% size error could increase the brightness temperature uncertainty by $\sim30\%$; 
however, this is not large enough to affect the main findings of our observations (in \S\ref{sec:results}) and discussions (in \S\ref{sec:discussions}), 
and therefore we do not discuss it further.

\subsection{Flux density upper limits for non-detected scans}

Even though significant fringes are not found for the long $>3D_{\rm Earth}$ baselines, it is possible to estimate their flux upper limits 
and obtain further insights about the most compact structure in the source (e.g., \citealt{johnson21}).
To this end, we analyzed more deeply the probability density distributions of fringes of all the scans with the SRT to determine at what $\sigma$
 level the space baselines are not detected.
For this analysis, we have primarily chosen PIMA because the software provides well-understood statistics for the fringe-fitting in the low S/N regime (see \citealt{petrov11}). We have also performed similar calculations with AIPS for cross-checks. 
The detailed procedures for the generation of the probability density distributions, their inspection, and cross-comparison between PIMA and AIPS results are discussed in Appendix \ref{appendix:pfd}.
By comparing the independent PIMA and AIPS fringe-fit results, we found consistent detections and non-detections, when a false fringe detection probability of $10^{-4}$ was chosen in PIMA (corresponding to PIMA and AIPS S/N threshold values of $\sim$6.1 and $\sim3.3$, respectively).
Accordingly, we ran the PIMA fringe search for all the scans and obtained the amplitude of each scan ($A_{\rm calib}$) or its upper limit ($A_{\rm upper}$), depending on their statistical significance with respect to the threshold false probability value $10^{-4}$.
Specifically, the amplitude $A_{\rm calib}$ is obtained by $A_{\rm calib}=A_{\rm raw}\times{\rm SEFD_{\rm net}}$ where $A_{\rm raw}$ is the raw visibility amplitude (i.e., correlation coefficients) and ${\rm SEFD_{\rm net}}=\sqrt{\rm SEFD_{1}SEFD_{2}}$ is the net system equivalent flux density that can be estimated from the antenna gains and system temperature information of two antennas 1 and 2 of each baseline.
As for the non-detection cases, $A_{\rm upper}$ is obtained by $A_{\rm upper}=A_{\rm raw}\times{\rm SEFD_{\rm net}}\times(SNR_{\rm det}/SNR)$ where $SNR$ is the S/N value from PIMA and $SNR_{\rm det}=6.1$ is the aforementioned detection threshold (i.e., $A_{\rm upper}=A_{\rm calib}$ if $SNR=6.1$ and $A_{\rm upper}>A_{\rm calib}$ otherwise).

Since this analysis can be extended to any dataset from {\it RadioAstron} observations of \m, we have also incorporated more interferometric 22\,GHz flux density measurements of \m from the {\it RadioAstron} AGN survey observations.
Applying the same analysis by PIMA yielded the corresponding flux upper limits on baseline lengths of up to $\sim25\,$G$\lambda$.

We additionally note on the uncertainties of $A_{\rm upper}$. We consider the S/N of each data point to be estimated as accurately as possible, based on the dedicated Monte Carlo simulations. Thus, the dominating error can originate from limitations in our knowledge of the station gain and system temperature information. 
Inferring their uncertainties from previous {\it RadioAstron} experiments (e.g., \citealt{gomez16,giovannini18,bruni20}), we consider the overall accuracy of the flux upper limit to be accurate on the order of $\sim10-20$\% at this radio frequency. 
However, further systematic uncertainties can arise due to, for example, antenna pointing errors, which can rather significantly change the value of $A_{\rm upper}$. While the antenna pointing errors can be largely corrected by the amplitude self-calibration in imaging observations, it is difficult to do so for the single baseline AGN survey.
We further note that those systematic effects tend to reduce the correlated amplitudes. Accordingly, we expect that the long space baseline $A_{\rm upper}$ values can be uncertain and larger by factors of $\sim2$ than what we estimated, if a conservative systematic pointing offset of half the beam is adopted.

\section{Results}\label{sec:results}

\subsection{RadioAstron 22\,GHz images}\label{subsec:kband_images}

\begin{figure}[ht!]
    \plotone{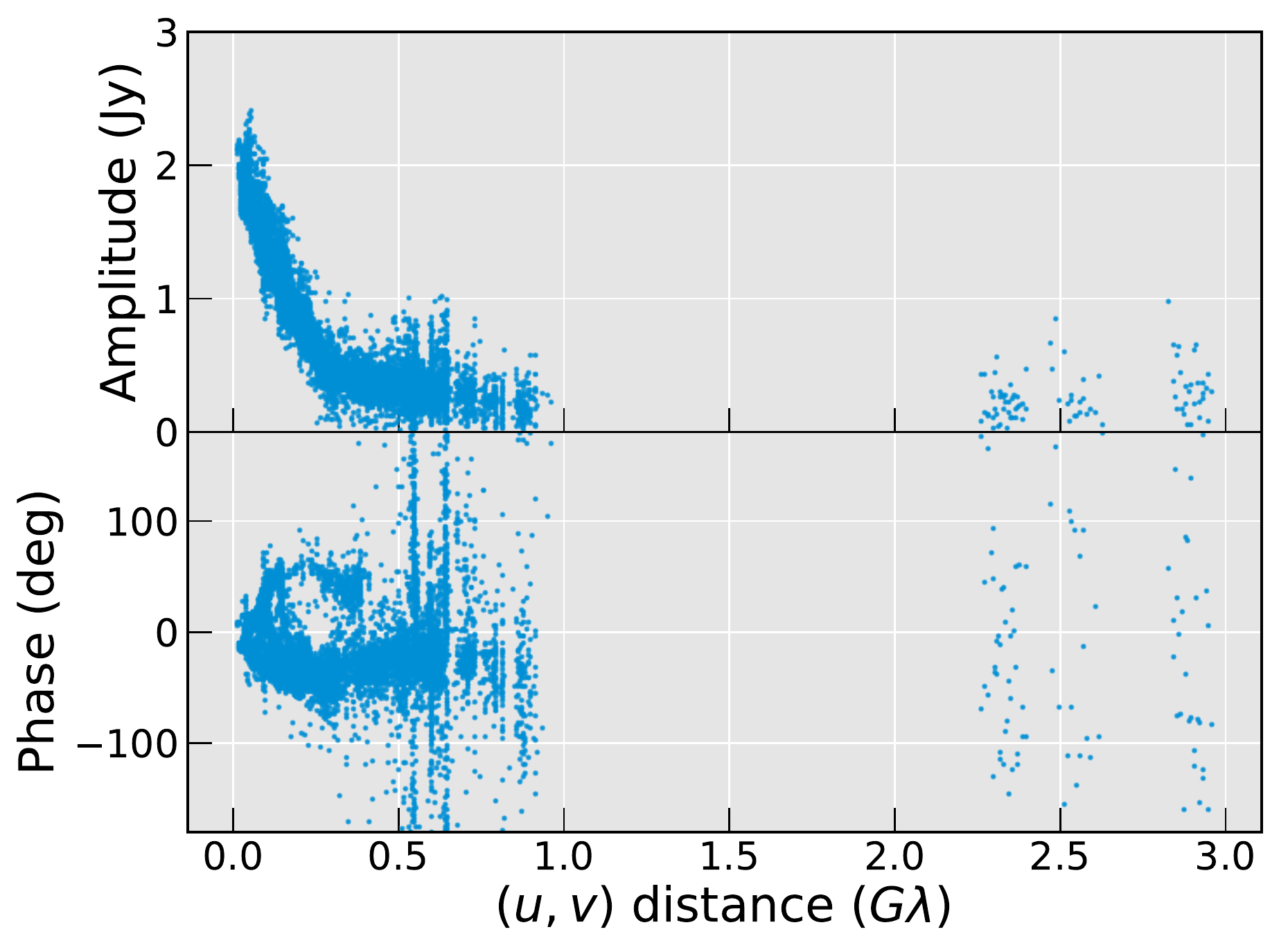}
    \caption{Correlated Stokes $LL$ flux densities and phases versus $(u,v)$-distance in the Fourier domain, after final amplitude and phase self-calibration. The data have been averaged over 60\,s for display.
    }
    \label{fig:amp_phs_uvdist}
\end{figure}

In Figure~\ref{fig:amp_phs_uvdist} we show the fully calibrated visibilities of \m observed by {\it RadioAstron} at 22\,GHz. 
Fringes are clearly detected for both ground and space baselines at $<1$G$\lambda$. Fringes for the space baselines up to $\lesssim3\,$G$\lambda$ are also detected, although at significantly decreased S/N.
We refer to Table \ref{tab:pfd} where the probabilities of false fringe detection for specific AIPS S/N values are estimated according to our simulation.
The large flux density of $\sim2$\,Jy at the shortest $(u,v)$-spacing decreases to $\sim250\,$mJy at $\sim2.2-3.0\,$G$\lambda$, indicating significantly resolved structure of the nucleus.

\begin{figure*}[ht!]
    \plottwo{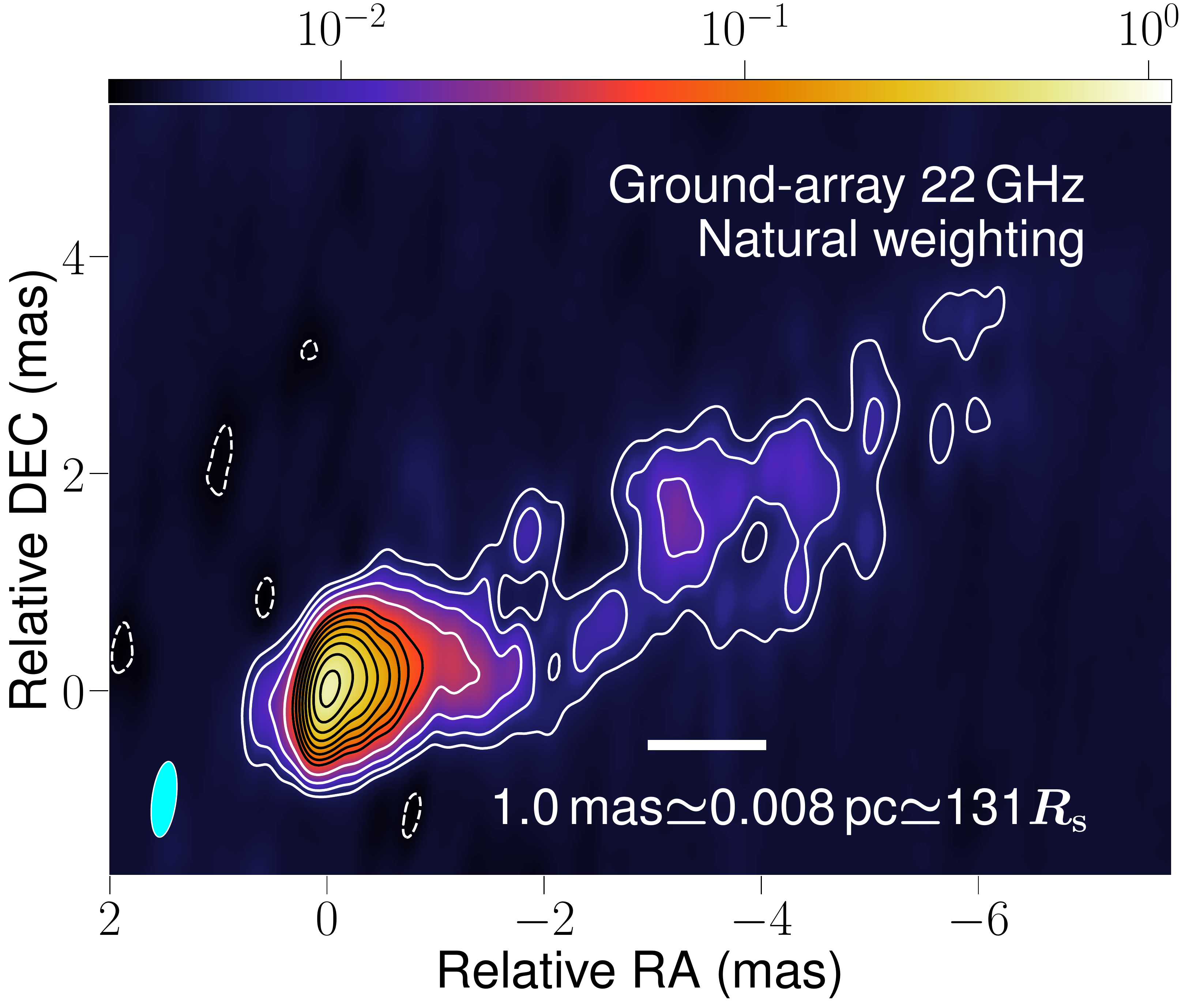}{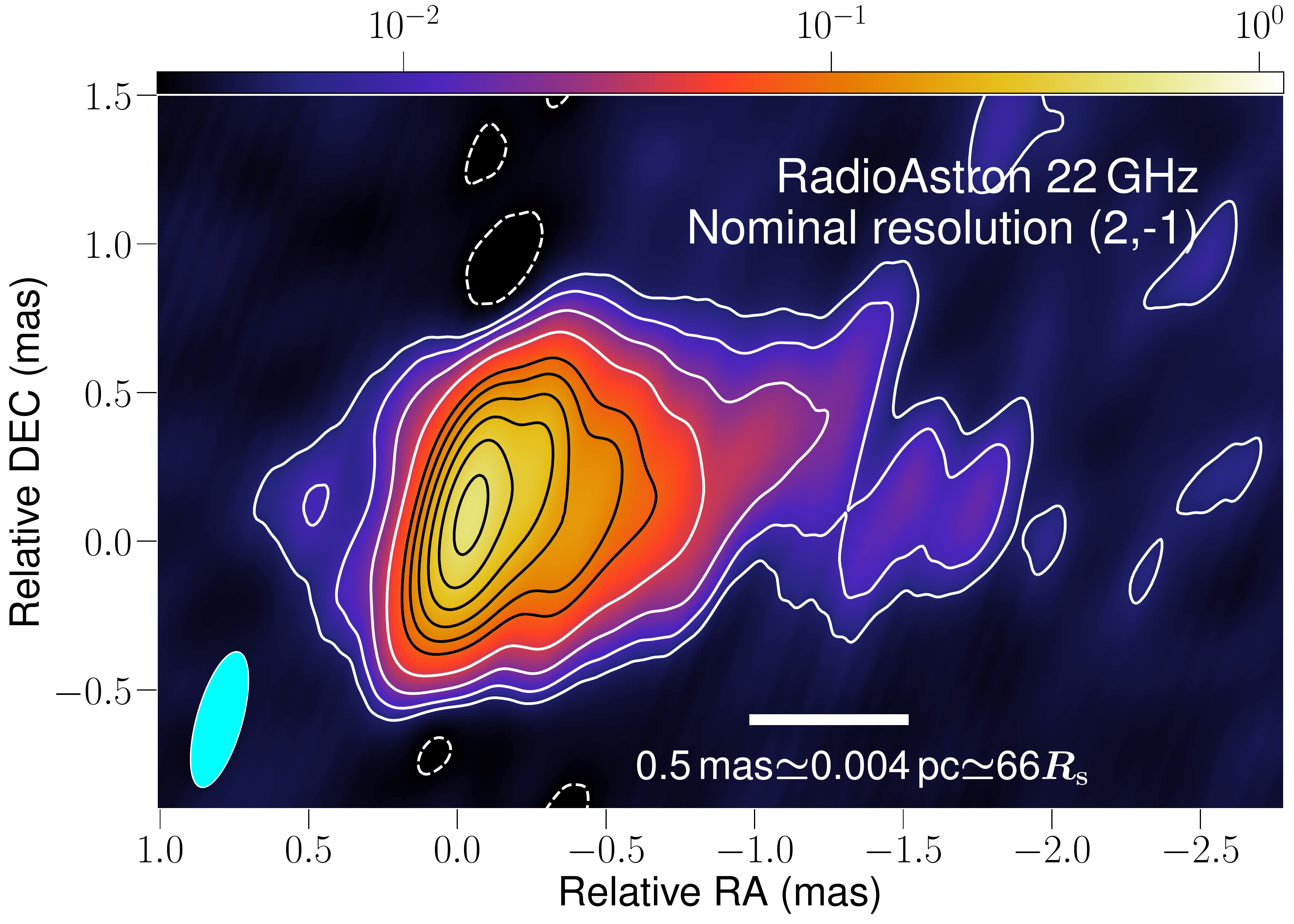}
    \plotone{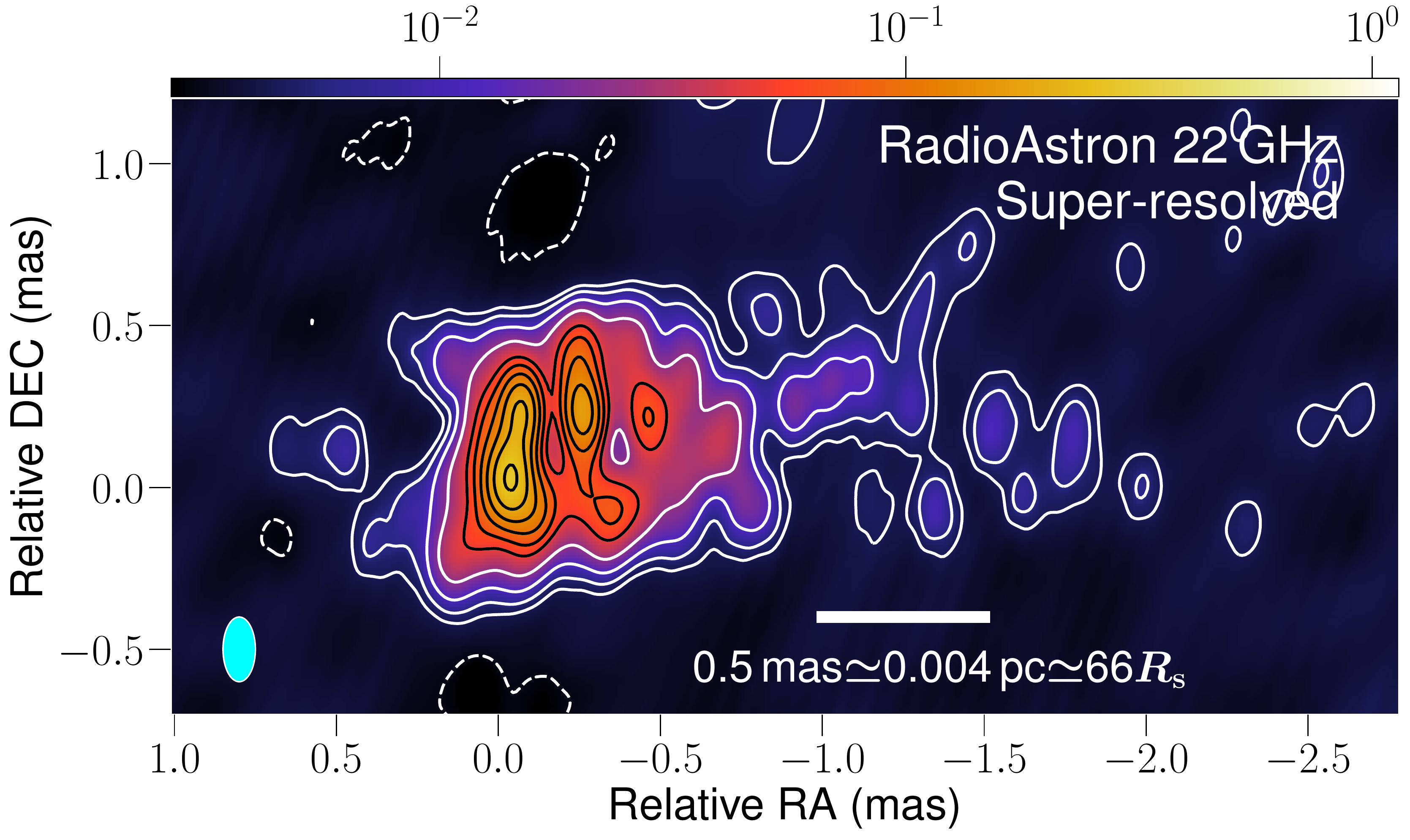}
    \caption{
    Images of the jet in \m at various angular resolutions.
    In all panels, color shows the total intensity in Jy/beam.
    Contours denote total intensity, starting from $\pm$3.5, 4.0, and 3.0 mJy/beam  respectively from top to bottom (dashed lines for negative), and increase in steps of $\sqrt{2}$.
    The colors of the contours are only used to increase clarity and do not encode physical meaning.
    Cyan ellipses at the left bottom corners of each panel indicate the interferometric beams.
    White ticks with text below show the angular and spatial scales in each image.
    {\it Top left:} ground array-only image with natural weighting, at a beam size of $0.7\times0.2$\,mas at a position angle of $-7.5^{\circ}$.
    {\it Top right:} space VLBI image with the mixed (2,-1) weighting (\S\ref{sec:ra_datareduction}), at a beam size of $0.47\times0.15$\,mas at a position angle of $-15.7^{\circ}$.
    {\it Bottom:} the same as above, but with slight super-resolution ($0.2\times0.1$\,mas at a position angle of $0^{\circ}$).
    }
    \label{fig:final_images}
\end{figure*}

In Figure~\ref{fig:final_images}, we also show the corresponding images of  \m at various angular resolutions.
The main parameters of those images, 
including the observing epoch, participating stations, observing frequency, total flux density, the image peak flux density as well as the image rms level, and achieved angular resolutions for specific visibility weighting,
are listed in Table~\ref{tab:image_parameters}.
The naturally weighted ground-only image consists of the bright compact core, extended jet which slightly bends to the north at $\sim2\,$mas core distance, and a weak counterjet. However, the limited angular resolution of the ground-only array, especially in the N-S direction, does not yet allow a clear view toward detailed structure of the jet at $\lesssim1\,$mas, such as limb-brightening. With the addition of {\it RadioAstron}, the jet becomes more clearly resolved both in the E-W and N-S directions. Applying slight over-resolution (comparable to the super-uniform weighting beam), the image reveals further edge-brightened jet and counterjet on $\sim0.3$\,mas distances from the core.
This morphology resembles the ``X-shaped'' inner jet structure (e.g., \citealt{kovalev07,walker18}; see Figures~22 and 23 of \citealt{walker18}) where both approaching and receding jets are edge-brightened, and thus supports the idea that the central engine of \m is located close to the VLBI core.

The shape of the core is also highly resolved with the slight over-resolution (Figure~\ref{fig:final_images} bottom), and its appearance is significantly elongated in the north-south direction, larger than the adopted beam size, with the brighter spot in the south.
Although such a structure was not seen in previous VLBI images of \m at 22\,GHz, we consider that this structure is real.
Specifically at this observing frequency, the angular resolution in the N-S direction of the new {\it RadioAstron} image is roughly $\sim0.45$\,mas 
for the (2,-1) weighting. This is significantly smaller than that of the previous VLBA observations \citep[cf.][]{hada13}, which is $\sim0.6\,$mas.
Also, we have parameterized the overall structure of the resolved nucleus by fitting an elliptical Gaussian model as described in \S \ref{subsec:modelfit} both in the Fourier and image domains, finding comparable values of the Gaussian parameters.
These numbers are summarized in Table \ref{tab:core_params}. 
Compared to the previous VLBI core size measurements for \m at 23\,GHz (see, e.g., Table\,1 in \citealt{hada13}), the new FWHM sizes reported here -- $\sim0.13\,$mas and $\sim0.36\,$mas along the minor and major axes, respectively -- are generally consistent with the previous measurements, except for the more highly resolved size in the E-W direction. 
The N-S elongation of the core of $\sim0.36$\,mas is also comparable to the angular resolution of the {\it RadioAstron} observation with uniform weighting. This further supports that the overall elongation and complicated substructure of the nucleus is resolved and real.

We also calculate the apparent brightness temperature, $T_{\rm B}$, of the core in the observer's frame using the core parameters in Table \ref{tab:core_params}, by
\begin{equation}
T_{\rm B}=1.22\times10^{12}\frac{S_{\rm core}}{\nu^{2}_{\rm core}\psi_{\rm maj}\psi_{\rm min}}(1+z)    
\label{eq:tb}
\end{equation}
\noindent \citep{kim18} where
$S_{\rm core}$ is the core flux density in Jy,
$\nu_{\rm core}$ is the observing frequency in GHz, 
$\psi_{\rm min, maj}$ are respectively the Gaussian FWHM component sizes along the minor and major axes in mas, and
$z=0.00436$ is the redshift of \m \citep{smith2000}. Using Eq. \ref{eq:tb} and parameters in Table \ref{tab:core_params}, we obtain $T_{\rm B}=(4.4\pm0.8)\times10^{10}$\,K.
This value is slightly lower than those from 15\,GHz VLBI observations of \m (e.g., \citealt{kovalev05}) and higher than those from measurements at $\gtrsim86\,$GHz (e.g., \citealt{hada16,kim18,eht2019}),
following the general trend of decreasing $T_{\rm B}$ of the VLBI core of various AGN 
at high radio frequencies (e.g., \citealt{lee08,kim18,nair19,cheng20}).

\begin{figure}[ht!]
    \plotone{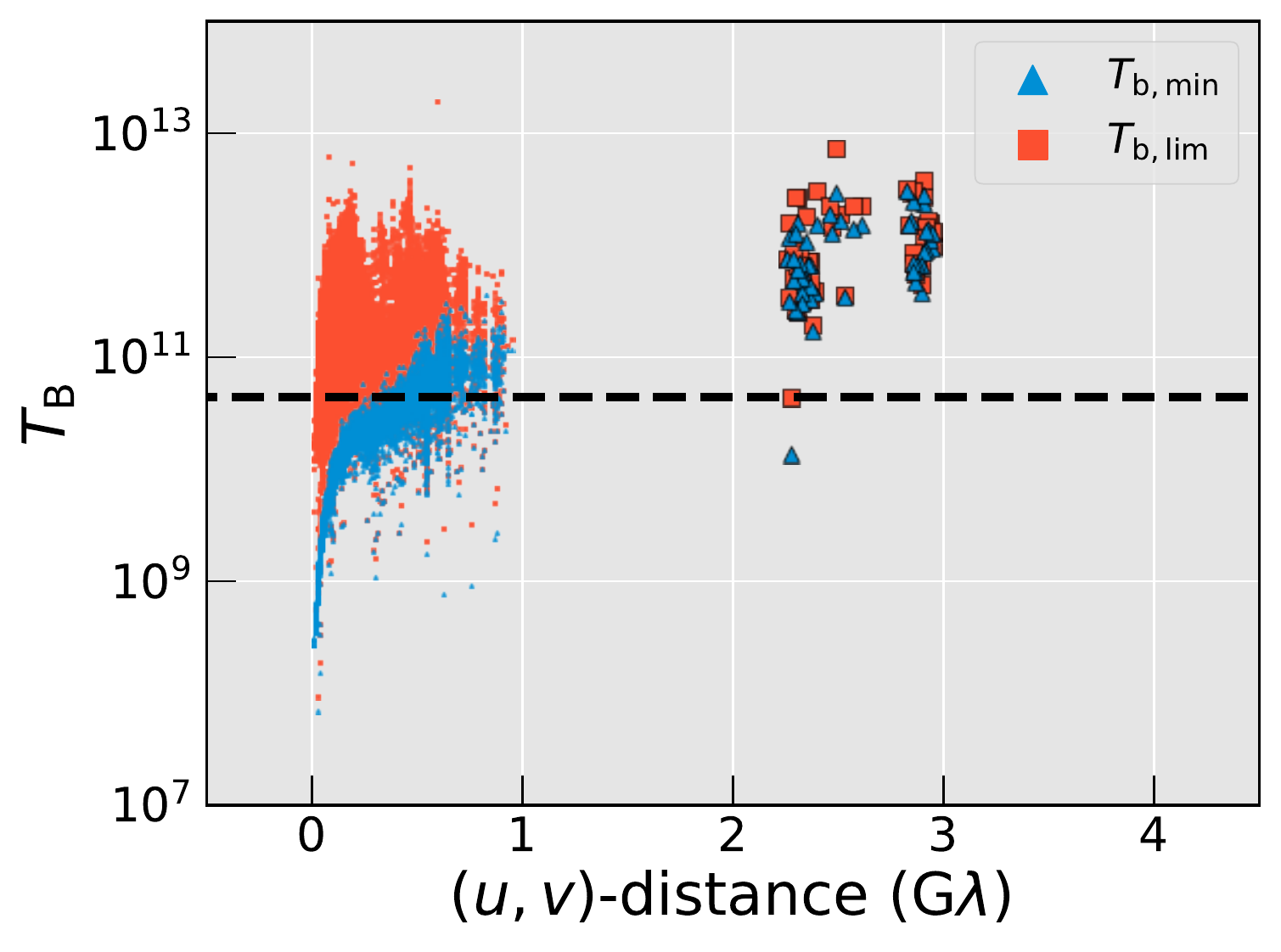}
    \plotone{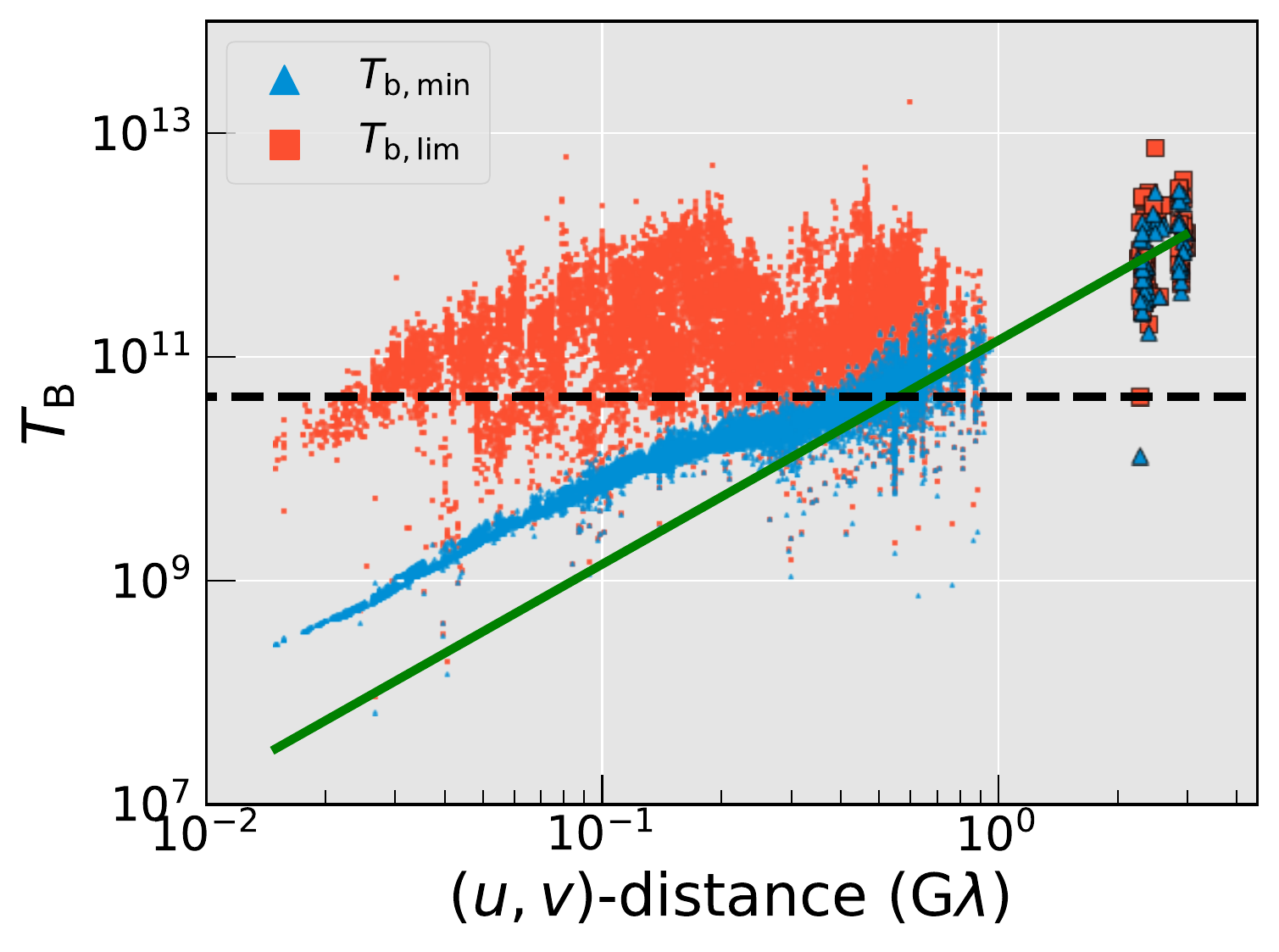}
    \caption{Minimum ($T_{\rm B,min}$; blue) and limiting ($T_{\rm B,lim}$; red) brightness temperature values calculated using visibility amplitudes and following \cite{lobanov15}. 
    The long ground-space baseline measurements (at $>1D_{\rm Earth}$) are highlighted by dark edges. 
    The black dashed horizontal line indicates $T_{\rm B}=4.4\times10^{10}\,$K from the Gaussian model-fitting.
    The green line in the bottom panel illustrates $T_{\rm B, min} \propto B^{2}$ dependence where $B$ is the baseline length (see the main text for details).
    Top and bottom panels show the same datasets but in different scales to illustrate the $T_{\rm B}\propto B^{2}$ dependence.
    }
    \label{fig:tb_lobanov}
\end{figure}

In addition, we also directly estimate lower limits on the brightness temperature of the source, following \cite{lobanov15} whose method uses only the correlated flux densities with less assumptions on the detailed model geometry, and thus suitable for experiments with sparse $(u,v)$-coverage. 
We compute the minimum ($T_{\rm min}$) and limiting ($T_{\rm lim}$) brightness temperature, which provides a range of actual brightness temperature (see \cite{lobanov15} for details). 
The results are shown in Figure~\ref{fig:tb_lobanov}, together with the estimate of $T_{\rm B}$ from Eq. \ref{eq:tb}.
As shown, the values of $T_{\rm B,min}$ agree well with the geometric modelfit result at $\sim$1\,G$\lambda$, while significantly larger $T_{\rm B,min}$ values appear at $\gtrsim$2\,G$\lambda$ (mean value $<T_{\rm B,min}>\sim10^{12}\,$K at $2-3\,$G$\lambda$). This is likely related to the compact substructure in the VLBI core, which is suggested by Figure~\ref{fig:final_images} bottom panel. 

We also calculate the diameter and opening angles of the limb-brightened inner jet and counterjet as follows.
First, we take the peak of the image as the center and circularly slice the jet side of the slightly over-resolved image (Figure~\ref{fig:final_images}, bottom) at varying radial distances from the center.
For each circular slice, we determine the positions of two local intensity maxima, to extract the ridge lines of the bright limbs.
The linear distance between the two local maxima at a specific radial core distance, $z$, is then used as the full width of the jet, $W$, and the apparent opening angle of the jet, $\phi_{\rm app}$, is calculated by  $\phi_{\rm app}=2\times\arctan(W/(2z))$ (e.g., \citealt{pushkarev17,kim18b}).
The same procedure was applied to the counterjet side.
For both jet and counterjet sides, the above calculations worked best at $0.2\,{\rm mas}\lesssim z \lesssim 0.5\,{\rm mas}$. 
At $z\lesssim 0.2\,{\rm mas}$, the bright core significantly contaminated the jet flux. 
At $z\gtrsim 0.5\,{\rm mas}$, the complicated jet structure and limited array sensitivity made it difficult to identify the edges of the limb-brightened jet. 
Therefore, we focused only on the reliable measurements which were made at $z\sim0.32-0.46\,$mas.
At this core distance, we obtain
$W_{\rm j}=0.44\pm0.10\,$mas ($W_{\rm cj}=0.57\pm0.10\,$mas) and 
$\phi_{\rm app, j}=48^{\circ}\pm11^{\circ}$ ($\phi_{\rm app, cj}=55^{\circ}\pm10^{\circ}$) 
where the subscripts j and cj denote the jet and counterjet, respectively.
Here, the uncertainties for the width, $\sigma_{W}$, are taken as 20\% size of the interferometric beam in the N-S direction, and the angle uncertainties, $\sigma_{\phi}$, are obtained by propagating the errors on $\sigma_{W}$.

The jet-to-counterjet brightness ratio (BR) is also estimated, using again the slightly over-resolved image. Since our single epoch observation does not provide the jet kinematics, we rely on the integrated flux densities of the jet and counterjet over certain extended regions, instead of trying to identify pairs of jet and counterjet components.
By integrating the jet and counterjet emission from 0.20\,mas to 0.45\,mas core distance along the jet at the position angle of $-72^{\circ}$ from north to east, we find $BR\sim10\pm3$. Here we adopt a characteristic systematic uncertainty $\sigma_{\rm BR}=3$, which represents variation of the value of $BR$ upon slight change in the shift of the integration region by, for example, $\sim0.05$\,mas. This BR value falls in the range of those from previous observations (see, e.g., \citealt{ly07,kovalev07,hada16,walker18,kim18}), however with some discrepancies, notably with \cite{hada16} who observed the object also in Feb 2014. This is further discussed in \S \ref{sec:discussions}.

\begin{table*}[]
    \centering
    \begin{tabular}{cccccccc}
        \hline
        Epoch & Array & $\nu$ & $S_{\rm tot}$ & Peak & $\sigma$ & Beam \\
        (YYYY-MM-DD) &   & (GHz)  & (Jy)     & (Jy/beam) & (mJy/beam) & (mas$\times$mas, deg) & {\sc uvweight} \\
         \hline
        2014-02-04 & SRT+BD+EF+GB+HH+KZ+SV+ & 22.236  & 2.19 & 0.52 & 1.0 & $0.47\times0.15$, $-15.7^{\circ}$ & (2,-1)  \\
         & TR+YS+ZC+VLBA(10) &  &  & 0.29 & 1.0 & $0.20\times0.10$, $0.0^{\circ}$ & (2,0)$^{a}$ \\
         2014-02-05 & VLBA(10) & 43.136 & 0.99 & 0.48 & 1.7 & $0.37\times0.14$, $-1.61^{\circ}$ & (2,0) \\
         \hline
    \end{tabular}
    \caption{
    Details of the images shown in \S\ref{sec:results} (Figures~\ref{fig:final_images} and \ref{fig:vlba_qband}). 
    From left, each column shows
    (1) the mean observing epoch in year-month-date format,
    (2) the observing array,
    (3) the central observing frequency,
    (4) the total VLBI flux density,
    (5) the image peak intensity,
    (6) the image rms noise, 
    (7) the beam size, and
    (8) the Difmap {\sc uvweight} parameter.
    (a) Approximately 63\% super-resolution in the N-S direction applied.
    }
    \label{tab:image_parameters}
\end{table*}

\subsection{Flux density upper limits at $>3$\,G$\lambda$ baselines}\label{subsec:ra_flux_upper_limits}

\begin{figure}[th!]
    \plotone{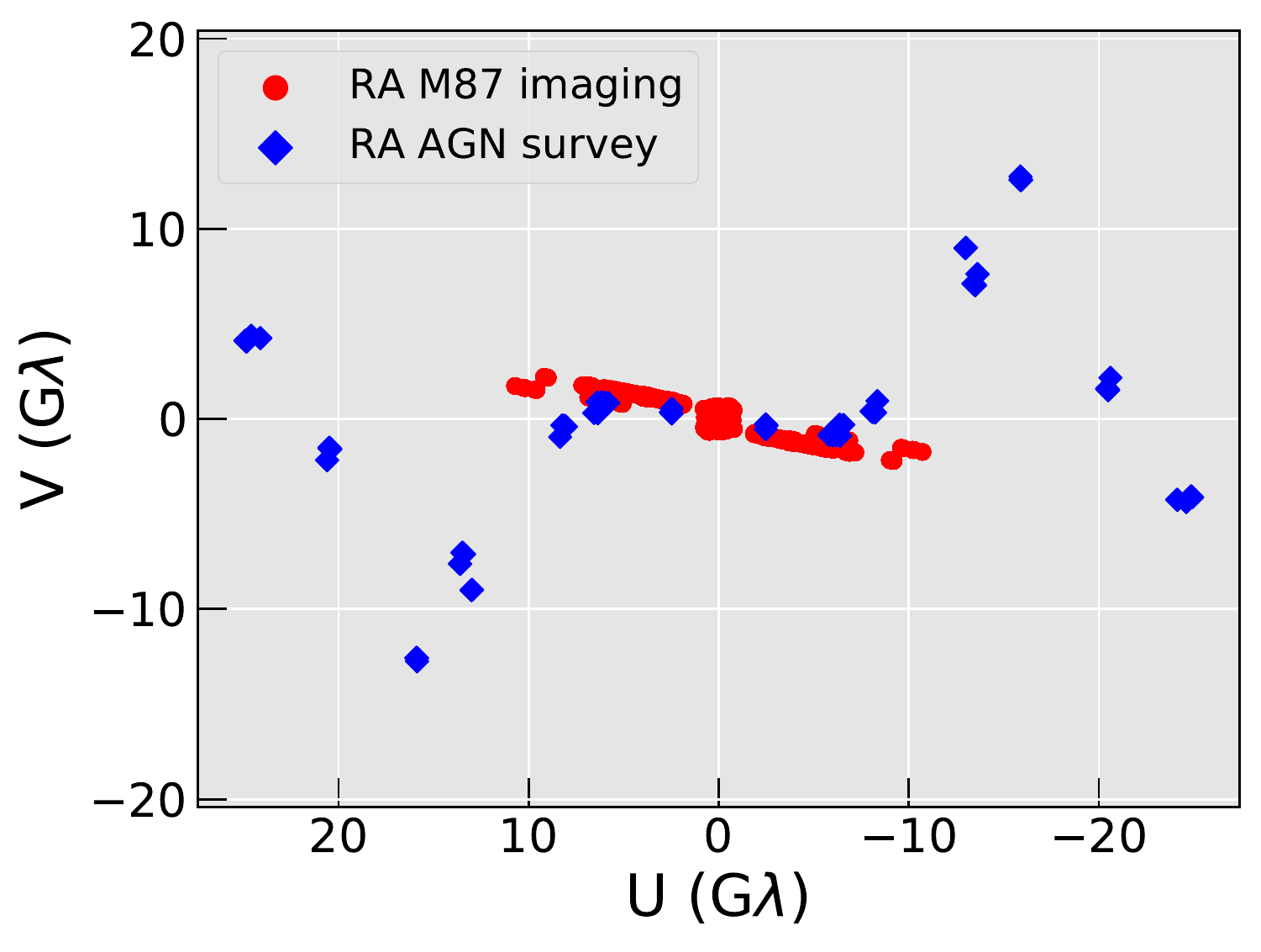}
    \caption{
    $(u,v)$-coverage of \m from the imaging and AGN survey observations.
    }
    \label{fig:survey_obs}
\end{figure}

Besides the images, we also show results of the further analysis of the ground-space fringe detections and the flux density upper limits,  including data from both the imaging and AGN survey observations (\S \ref{subsec:ra_survey_data}). The combined $(u,v)$-coverage is shown in Figure~\ref{fig:survey_obs}. There is no clear fringe detection above the PIMA S/N threshold of 6.1 at $\sim3-25$G$\lambda$ baseline lengths. Accordingly, we have only calculated the flux density upper limits. All the detailed information including observing epoch, experiment code, two-letter code for the ground station, solution interval, values of S/N of actual data and detection threshold, length and position angle of the baseline, and derived flux upper limits are summarized in Table \ref{table:flux_upper_limits} in Appendix B. 
Figure~\ref{fig:survey_obs_flux} also shows summaries of the flux densities and the upper limits by combining both the imaging and survey observations.
At the short $(u,v)$ distances of $\sim(0.2-0.8)$G$\lambda$, flux densities derived from the full AIPS data reduction and Difmap imaging with self-calibration agree well with the independent flux estimates from the PIMA analyzes (middle panel of Figure~\ref{fig:survey_obs_flux}).
This provides 
confidence that
the choice of the PIMA S/N threshold and a-priori amplitude calibration of the SRT are valid.
Keeping this in mind, we note that the PIMA analysis indicate significantly small flux densities of $\lesssim100$\,mJy at $\sim3-25\,$G$\lambda$, with the most tight constraint given by 
the phased JVLA to SRT baselines at $\sim15.8$\,G$\lambda$ with an upper limit of $\sim70$\,mJy upper limit (Table \ref{table:flux_upper_limits}).
Also, we emphasize that these non-detections were presented in  various epochs during 2013--2016 as well as different baseline position angles. This indicates that the jet of \m at 22\,GHz persistently lacks compact features that could be detected by the long space baselines.

\begin{figure}[th!]
    \plotone{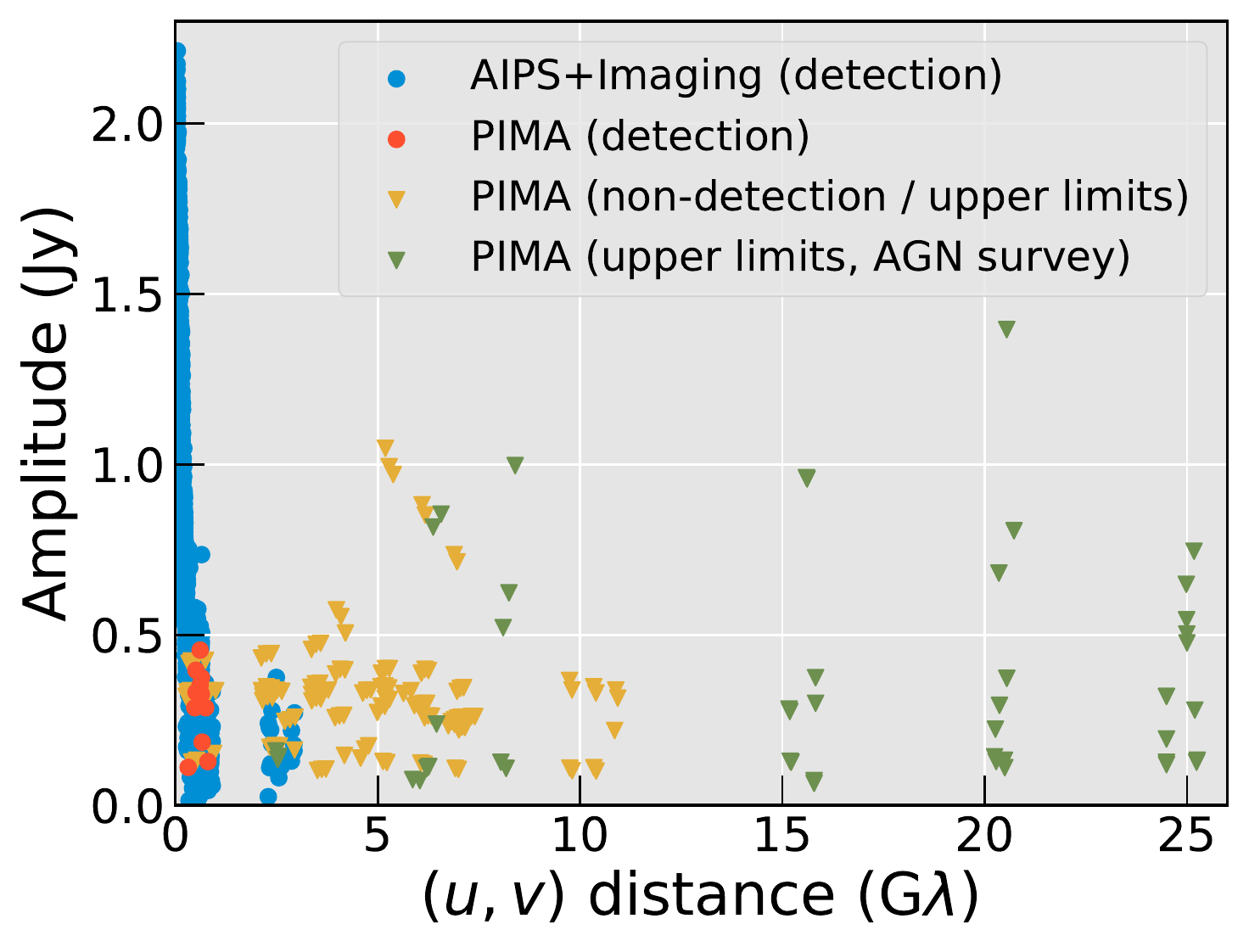}
    \plotone{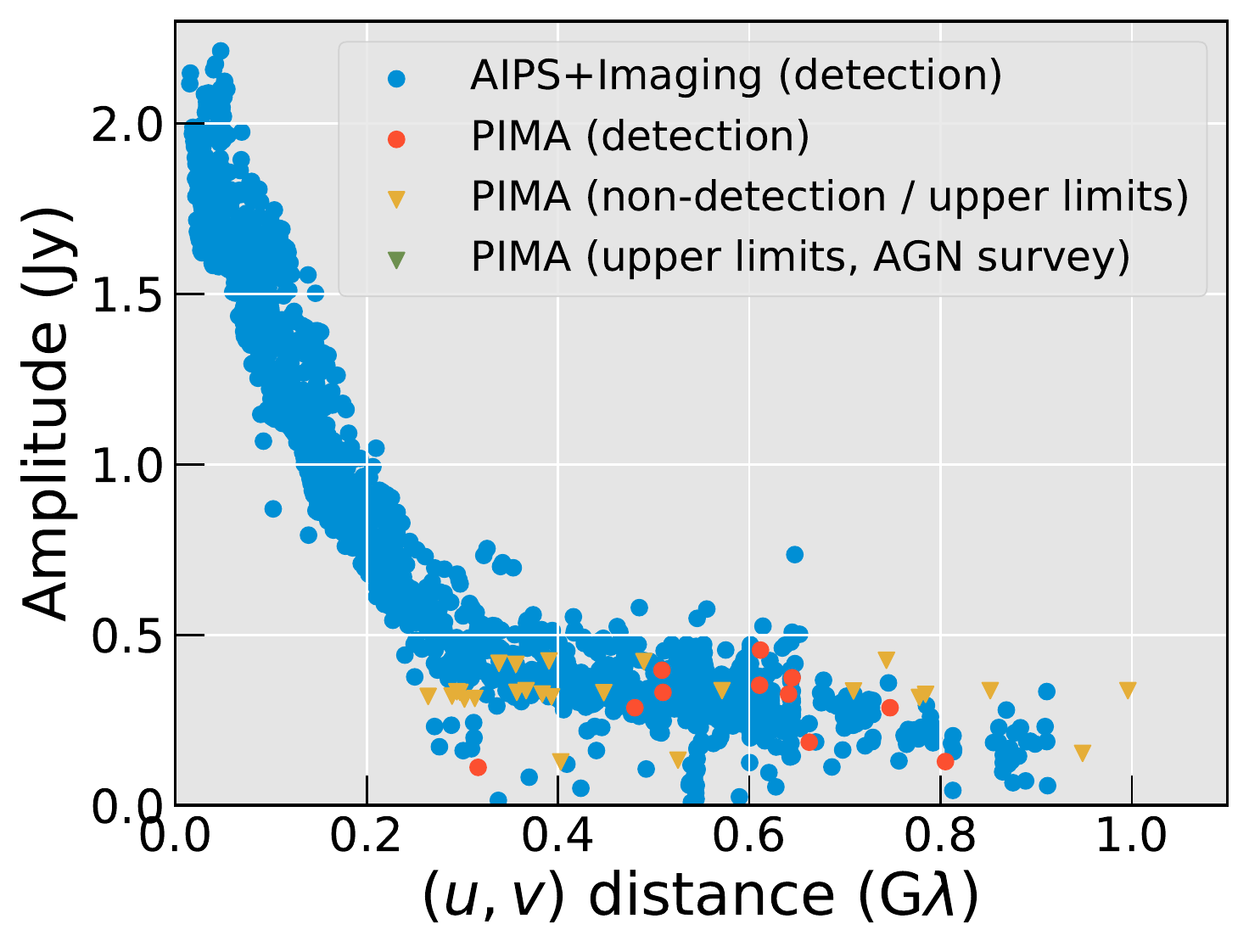}
    \plotone{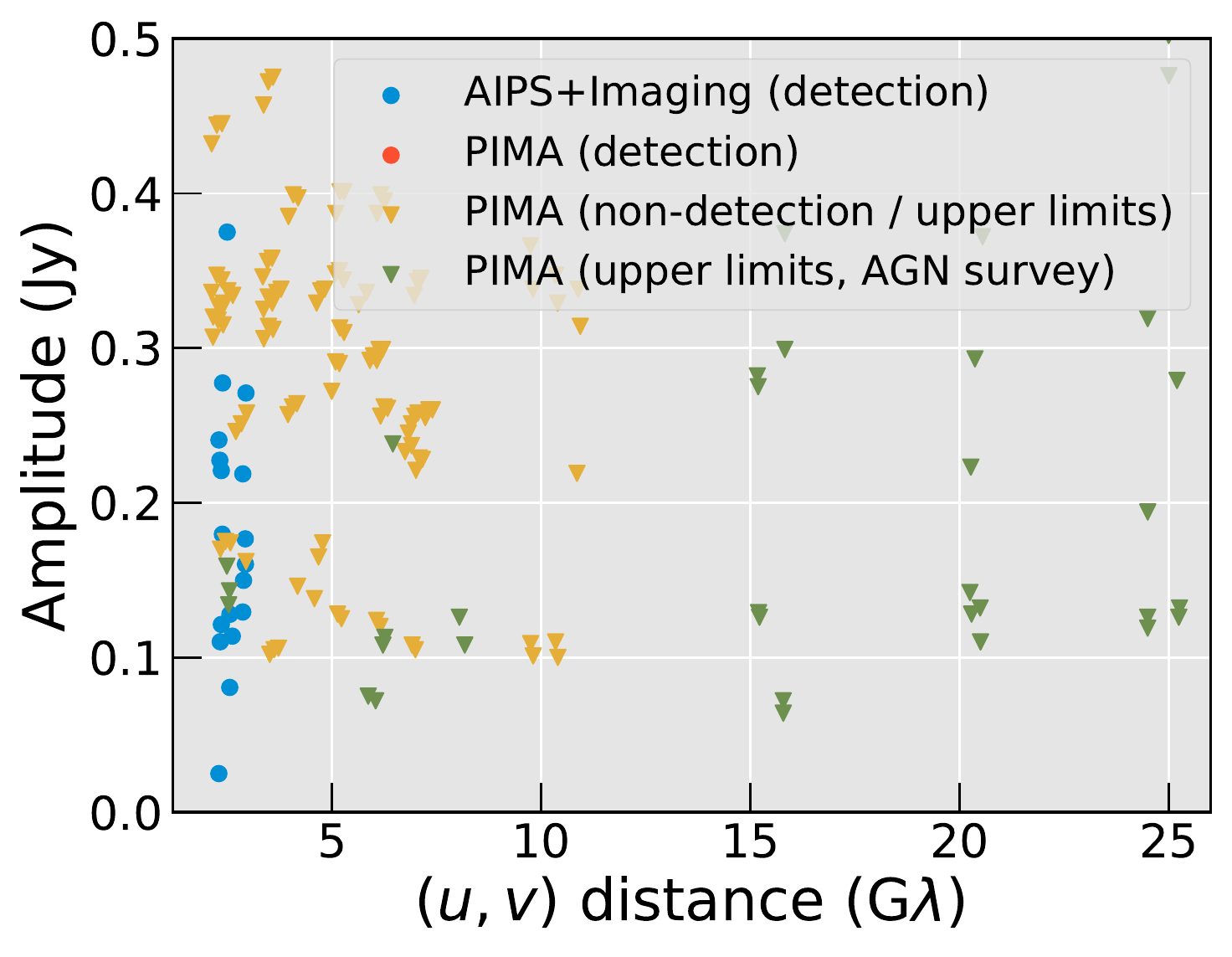}
    \caption{Flux densities and their upper limits of \m at 22\,GHz from the {\it RadioAstron} imaging and AGN survey observations, for 
    ({\it top}) the entire $(u,v)$-coverage, 
    ({\it middle}) the ground spacings, and
    ({\it bottom}) the space baselines.
    The PIMA measurements and upper limit estimations were made with a solution interval of 570\,s.
    The AIPS+imaging dataset have also been averaged over the same timescale.
    We note noticeable flux density dropouts at $\sim0.2-0.6$\,G$\lambda$ baselines, which are due to the intrinsic source structure and time-smearing effects.
    }
    \label{fig:survey_obs_flux}
\end{figure}

\subsection{VLBA 43\,GHz image}

\begin{figure}[th!]
    \plotone{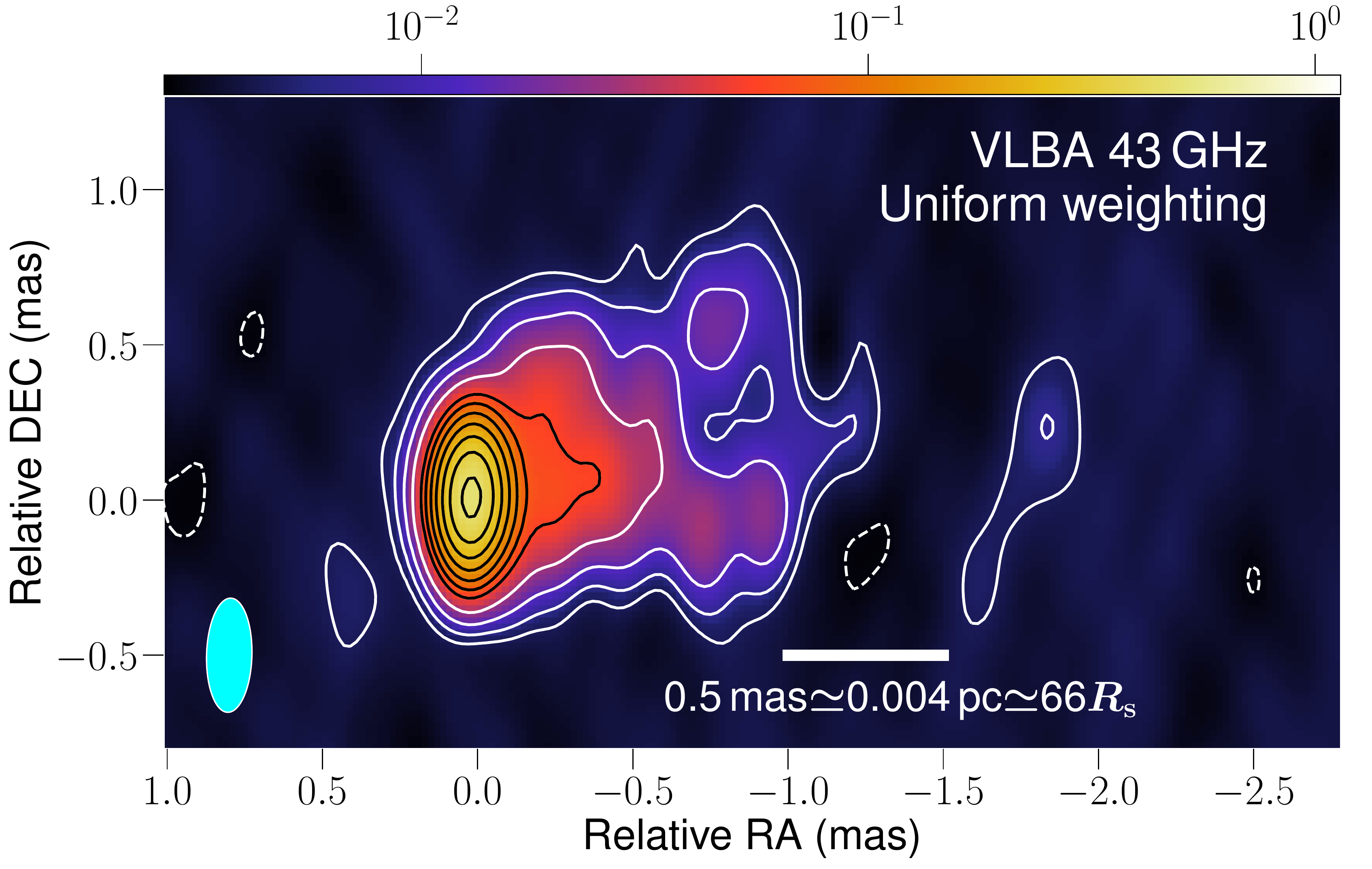}
    \caption{The same as Figure~\ref{fig:final_images} but for the VLBA 43\,GHz, with a beam of $0.37\times0.15$\,mas at a position angle of $-1.61^{\circ}$ from the uniform weighting.
    Again, we note that colors for the contours are only for increasing clarity.
    }
    \label{fig:vlba_qband}
\end{figure}

Figure~\ref{fig:vlba_qband} shows the image of \m from the quasi-simultaneous VLBA observations at 43\,GHz, made using uniform weighting (because of higher S/N fringe detections at the long baselines than at 22\,GHz). 
Detailed properties of this image are given in Table \ref{tab:image_parameters}. 
The 43\,GHz image shows the compact core and  edge-brightened approaching jet, similar to typical VLBA 43\,GHz maps of \m over the last decade \citep{walker18} and the one obtained on 2014 Mar 26 (see Figure~6 of \citealt{hada16}), which is close in time to our observation (49\,days later). 
Our 43\,GHz image does not clearly reveal the counterjet emission (\S\ref{subsec:kband_images}), which can be due to a combined effect of the limited image dynamic range as well as fainter jet emission at the higher observing frequency.
Therefore, we set a lower limit on the jet to counterjet BR by following the same BR calculations in \S\ref{subsec:kband_images} and at the 0.20--0.45\,mas core distances. From this, we obtain BR$>$6 at 43\,GHz. If there is no frequency dependence for the jet to counterjet BR, we can combine the results at 22 and 43\,GHz, which yields BR\,$\sim6-11$. This value is consistent with results of \cite{hada16} who reported BR\,$\sim5-20$ at the 43--86\,GHz frequency band and within 1\,mas from the core in the same 2014 Feb epoch, as well as \cite{kim18} who also suggest BR\,$\lesssim20$ within $\sim0.3$\,mas from the core at 86\,GHz.

Since the 22 and 43\,GHz observations are interleaved to each other, we can compute the spectrum of the jet in \m with negligible time-variability effect.
A more comprehensive analysis of the jet spectrum, also including the 4.8\,GHz {\it RadioAstron} image (Kravchenko et al. in prep), can be done with precise alignment of the multi-frequency jet images and is planned for a future work. 
In this work, we instead provide the integrated spectrum of the core for later discussions.
The spectral index of the nuclear region is estimated by 
matching the angular resolutions of the 22 and 43\,GHz maps and computing the ratio of the peak intensities by $\alpha=\log(I_{43}/I_{22})/\log(43/22)$.
For this, we adopt for both images a common convolving beam of 
$0.47\times0.15\,$mas at $0^{\circ}$ position angle and obtain the peak intensities of 0.52 and 0.49 mJy/beam at 22 and 43\,GHz, respectively.
Assuming $\sim10\%$ absolute flux calibration errors, the core spectral index $\alpha=-(0.1\pm0.2)$. This result is consistent with earlier studies of the partially optically thick nucleus at $\lesssim43\,$GHz (e.g., \citealt{ly07,hada12,kim18b,kravchenko20b}).

\begin{table*}[t]
    \centering
    \begin{tabular}{cccccc}
    \hline
    Method & $S_{\rm core}$ & $\psi_{\rm min}$ & $\psi_{\rm maj}$ & $\rm PA_{\rm core}$ & $T_{\rm B}$ \\
    (1) & (2) & (3) & (4) & (5) & (6) \\
        & (Jy)  & (mas) & (mas) & (deg) & ($10^{10}\,K$) \\
    \hline
    Difmap & 0.74 & 0.12 & 0.32 & $-3.38$ &  \\
    JMFIT & 0.91 & 0.13 & 0.40 & $-3.60$ &  \\
Nominal  & $0.83\pm0.09$ & $0.13\pm0.01$ & $0.36\pm0.04$ & $-(3.49\pm0.11)$ & $4.4\pm0.8$ \\
    \hline
    \end{tabular}
    \caption{
    Parameters of the core.
    (1) Methods used for the Gaussian model-fitting of the core. Difmap and JMFIT respectively indicate fitting methods working in the visibility and image domains, respectively. The nominal values and their uncertainties are obtained by averaging and half the difference of the two independent estimates.
    (2) Flux density of the model-fitted core,
    (3) and (4) lengths of the minor and major axes of the Gaussian, respectively.
    (5) Position angle of the major axis as measured from north to east, and
    (6) Apparent brightness temperature of the core.
    }
    \label{tab:core_params}
\end{table*}

\section{Discussion}\label{sec:discussions}

One immediate result of the {\it RadioAstron} 22\,GHz observations of \m, both from the imaging and AGN survey experiments, is that the fringe detection is limited only up to $\sim3\,\mathrm{G\lambda}$. On the other hand, ground-based observations of \m at significantly shorter wavelengths, especially at 1.3\,mm by the EHT, detect the source up to much longer $\sim$8\,G$\lambda$ \citep{eht2019}, with less dominance from the jet (thus different source responses at long baselines).
Although previous studies of \m reported the frequency-size dependence of the VLBI core and possible interpretations \citep[e.g.,][]{hada13}, the {\it RadioAstron} results provide much stronger evidence that the intrinsic size of the core of \m is significantly larger at longer wavelengths. The most reasonable explanation is given by the classical model of relativistic jet from \cite{bk79}, in which the photosphere of the jet moves outward from the central engine as the observing wavelength increases, due to progressively increasing synchrotron opacity of the jet plasma. This is well supported by the flat or inverted radio spectrum of the nucleus of \m at $\lesssim130\,$GHz \citep[see][]{kim18} and the apparent change of the core position at different wavelengths \citep{hada11}.

Although the above interpretation can explain the overall size of the whole VLBI core region, we note that the new space VLBI image also features highly resolved structure inside the ground-VLBI core (see Figure~\ref{fig:final_images}), at the angular resolution of $\sim370\times170\,\mu$as, which is the record-high value at the 22\,GHz frequency.
This is perhaps better revealed by the ground-to-space fringe detection and accordingly high brightness temperature of reaching up to nearly $T_{\rm B,min}\sim10^{12}\,$K. 
In fact, this $T_{\rm B,min}$ is one of the highest values for \m reported in the literature, approaching nearly the inverse-Compton catastrophe limit ($\sim10^{12}\,$K; \citealt{kellermann69}). For comparison, decadal VLBI monitoring of \m during 1994--2019 at comparable radio frequency of 15\,GHz, only using ground facilities \citep{homan21}, reports significantly lower values of  $T_{\rm B}\sim10^{10}-10^{11}\,$K. 
We note that the maximum brightness temperature detectable by interferometric technique depends on the physical baseline lengths (see, e.g., \citealt{lobanov15}), and thus {\it RadioAstron} can detect higher brightness temperatures than ground-only arrays could measure.
Also, the GMVA and EHT observe $T_{\rm B}\sim10^{10}\,$K in \m at 86 and 230\,GHz bands respectively, at even higher angular resolution of $\sim50$ and down to $20\,\mu$as (see \citealt{kim18b}). We note that the jet of \m and other surrounding emitting materials can be optically thin at 230\,GHz \citep{eht2019}. Thus {\it RadioAstron} and the EHT may look at physically different regions or depths in the jet.
Therefore, we aim to understand the high $T_{\rm B,min}$ with various physical scenarios.

To begin with, we compare the $T_{\rm B, min}$ estimate with the intrinsic brightness temperature, $T_{\rm B,int}$.
The exact value of $T_{\rm B,int}$ depends on the detailed physics of the jet plasma and can often be difficult to determine accurately. Nevertheless, we note that \cite{homan21} find typical $T_{\rm B,int}=(4.1\pm0.6)\times10^{10}$\,K at 15\,GHz in the milliarcsecond-scale cores of more than 100 AGN jets, especially when the jets were in their median state of activity.
This value is close to or slightly lower than the brightness temperature at the 
energy equipartition for the magnetic fields and emitting particles (i.e., electrons and positrons except for a pure proton-synchrotron jet), $T_{\rm B}\sim5\times10^{10}\,$K \citep{readhead94}.
Therefore, under the reasonable assumption that $T_{\rm B,int}=T_{\rm B,eq}$ at 22\,GHz,
the observed high $T_{\rm B, obs}$ could have been Doppler-boosted by the Doppler factor $\delta=T_{\rm B, obs}/T_{\rm B,int}\times(1+z)\gtrsim20$ where the last inequality is obtained by replacing  $T_{\rm B, obs}$ by $T_{\rm B, lim}\sim10^{12}\,$K. 
\footnote{
We note that the compact radio cores of blazars and \m differ in their nature because the former may represent a recollimation shock or a surface where the opacity is close to unity ($\tau=1$), while the latter could be plasma near the jet base (see, e.g., \citealt{hada11}). If the base of a jet is less kinetic energy dominated (e.g., \citealt{blandford19}), lower values of $T_{\rm B,int}$ are physically expected for the core of \m. This can raise the required values of $\delta$ higher.
}

We note that large values of $\delta\gtrsim10-20$ are common in sources such as blazars whose jets are almost directly pointed to the observer (e.g., within a few degrees offset; \citealt{hovatta09,homan21,jorstad17,weaver22,liodakis21}).
In contrast, \m has significantly larger viewing angle of $\sim15^{\circ}-30^{\circ}$ \citep{hada16,mertens16,walker18}, which in combination with measurable jet-to-counterjet brightness ratio 
gives significantly smaller values of $\delta\sim1-2$ \citep{kim18} within $\lesssim1\,$mas from the core.
More specifically, \cite{hada16} measured the \m jet kinematics in Feb-May 2014, when our {\it RadioAstron} observations were made.
The authors reported slow apparent speeds of $\sim\beta_{\rm app}\sim0.3-0.5$ within $\lesssim1\,$mas from the VLBI core. Thus, we can compute $\delta=(\Gamma(1-\beta\cos\theta))^{-1}$ where $\Gamma=(1-\beta^{2})^{-1/2}$ and $\beta$ are respectively the bulk Lorentz factor and intrinsic speed, finding again $\delta$ of only $\lesssim2$ assuming a relatively large jet viewing angle of $\sim15-30^{\circ}$. 
In the following, therefore, we examine in more detail alternative ways that can produce large Doppler factors in \m.

One way of substantially increasing the value of $\delta$ is to invoke more complicated jet geometry where the local viewing angle of the emitting component can be significantly smaller. For example, $\delta$ of the order of $\sim10$ can be achieved if $\theta\sim0.35^{\circ}$ for $\beta_{\rm app}\sim0.3$. 
This condition may be realized in the base of a jet where the jet opening angle can often be large or the jet internally rotates at significant speed (e.g., \citealt{eht2019}) and therefore a small blob of moving jet plasma inside the whole jet and line of sight can intersect by chance (e.g., \citealt{lenain08}).
We note that the intrinsic opening angle of the jet is as wide as $\sim60^{\circ}$ \citep{kim18}, which can offer the necessary geometry for the above scenario to occur.
In addition, the jet in \m shows a time-changing apparent position angle on a decadal timescale \citep{walker18}. 
The base of such a swinging jet can intersect with the line of sight and increase the beaming effect in certain periods.
More specifically, \cite{walker18} show a transverse shift of the jet ridge line of $\sim0.3\,$mas from the average position at $\,2$\,mas from the core. This can be translated into $\sim8.5^{\circ}$ position angle shift of the jet. If $\theta$ can increase or decrease by a comparable amount in certain periods, the chance for the  substructure inside a broad jet to intersect with the line of sight can be higher.

Another way of increasing $\delta$ is to have jet plasma whose true speed, that is responsible for the Doppler boosting of the observed emission, is much higher than observed (thus higher $\beta$ and $\Gamma$). 
This possibility has been suggested for \m based on the relatively large jet-to-counterjet brightness ratio versus often observed slow apparent motions in the source (e.g., \citealt{kovalev07}). In this regard, we note that the jet in \m is uniquely edge-brightened, indicating jet-transverse structures (see references in \citealt{kim18}). 
One of the models explaining the edge-brightened morphology is the so-called ``spine-sheath'' model (see, e.g., \citealt{pelletier89,macdonald15,macdonald17} and more) in which a highly relativistic, fast-moving beam (spine) is present inside a slower and broader wind (sheath).
To explain the edge-brightened morphology purely by the boosting effects, however, the inner spine should move faster than a critical speed so that $\delta$ becomes lower in the spine than the sheath; for example, $\delta\sim1.2-1.4$ in the spine versus $\delta\sim1.6-2.0$ in the sheath to explain the jet transverse brightness distribution near the VLBI core \citep{kim18}.
Therefore, the velocity stratification alone would not provide the necessary large Doppler factor of $\delta\gtrsim20$.

Besides the two simple cases, we remark that recent plasma physical models including detailed physics of turbulence and magnetic reconnection show the generation of
extremely energetic cells of plasma which can be ejected inside the jet towards observer (``jet-in-jet'';  see, e.g., discussions and references in \citealt{kim20}).
When such extreme ejection occurs, the necessary combination of smaller $\theta$ and larger $\beta$ than observed so far in \m may be locally realized. However, a detailed and more quantitative comparison of such models with our observations is beyond the scope of our paper.

In the above discussions, we have assumed $T_{\rm B,int}\sim5\times10^{10}\,$K.
On the other hand, \cite{liodakis21} show based on flares detected by long-term 15\,GHz flux monitoring of blazars that $T_{\rm B,int}$ can be as high as $\sim2.8\times10^{11}\,$K when the jets are in their active states. If we assume $T_{\rm B,int}=2.8\times10^{11}\,$K,  
 a significantly lower bound of $\delta\gtrsim3.6$ is obtained, significantly reducing the required $\delta$ in \m.
Below we list additional observational evidence supporting the presence of such high $T_{\rm B,int}$ in the VLBI cores of jetted AGNs.
First and most notably, {\it RadioAstron} observations of highly powerful jets in blazars and quasars 
(e.g., 
0836+710, \citealt{vega-garcia20};
3C\,273, \citealt{kovalev16};
BL Lac, \citealt{gomez16};
OJ\,287, \citealt{gomez22};
0716+714, \citealt{kravchenko20a})
find $T_{\rm B,min}\sim10^{13}\,K$, which, when reconciled with the Doppler factors from the VLBI kinematics, suggest $T_{\rm B,int}\gtrsim 10^{12}\,$K.
Such high $T_{\rm B,int}$ is also observed by ground-VLBI when the VLBI cores exhibit flares during ejection of a new VLBI component, as discovered also by \cite{jorstad17}.
In this regard, a remarkable example demonstrating that high $T_{\rm B,int}$ is a rather time-dependent state is the previous {\it RadioAstron} observations of 3C\,273 that revealed strong time variability of the observed $T_{\rm B}$ by two orders of magnitudes over multiple years ($\sim10^{13}$ to $\sim10^{11}$K; see \citealt{kovalev16,bruni16}).

From the physical point of view, the high $T_{\rm B,int}$, especially close to the inverse Compton limit, can be due to an increased number of energetic and emitting electrons,
for instance due to strong particle acceleration  by turbulence (e.g., \citealt{marscher14}) or other mechanisms (see discussions in \citealt{kovalev16}). 
The changing particle density can also cause significant variations in the nuclear opacity, as observed by time-variable shift of the VLBI core positions \citep{plavin19,chamani22}.
We note that the arcsecond-scale core of \m shows elevated X-ray flux density around 2014 \citep{sun18}, which could be supporting evidence for higher $T_{\rm B,int}$ in Feb 2014 than other epochs.

We then briefly discuss if $T_{\rm B,min}$ of $\sim10^{12}$\,K can be due to $T_{\rm B,int}$ as high as $\sim10^{11}$\,K, at least temporarily, in \m. We note that the AGN survey observations of the source by {\it RadioAstron} did not reveal significant fringes to the long space baselines over multiple years. 
Therefore, the fringe detection to the space and high $T_{\rm B,min}$ in 2014 may be interpreted as due to short time variability in the core of \m  (see, e.g., \citealt{acciari09,abramowski12}).
However, care is required for this interpretation.
We remark that existing VLBI light curves of the radio core of \m at 15, 22, and 43\,GHz (see \citealt{lister18,kim18b,walker18}) do not strongly support the particularly elevated flux state of the source in 2014. 
Also, the sensitivities of the {\it RadioAstron} imaging and AGN survey observations can differ, with the former being more sensitive due to the stacked baselines of ground stations upon successful fringe detections.
This makes a direct comparison of results from the imaging and survey observations sophisticated.
Regardless, we note that characteristic cooling timescale of a plasma with $T_{\rm B}\sim10^{12}\,$K is as short as $<10^{-4}\,$yr $\sim1$\,h at 22\,GHz according to \cite{readhead94}. 
Therefore, only continuous injection of energetic particles could have maintained the high $T_{\rm B}$ of \m during the {\it RadioAstron} observation over $\sim1$ day timescale.
The core of \m is located close to the central black hole (within subparsec scales; \citealt{hada11}), unlike typical blazars whose cores are thought to be located at least parsecs downstream of the black holes at centimeter wavelengths (e.g., \citealt{pushkarev12}).
After these considerations, the observed high $T_{\rm B,min}$ in \m does not exclude the possibility of intrinsically high brightness temperature, and thus strong on-site particle acceleration or injection of more energetic emitting electrons, in the base of a relatively low-power jet of \m.

In summary, we can conclude that it is not impossible that such a high $T_{\rm B}$ could exist in \m,
especially when the effects of both the Doppler boosting and high intrinsic brightness temperature are present.
However, constraining further the aforementioned models solely from the single epoch result is challenging.
Fortunately, \m was again observed by {\it RadioAstron} in 2018 at 22\,GHz. Thanks to the similar ground $(u,v)$-coverage, this dataset has the potential to confirm the ground-space fringe detection at 22\,GHz and constrain the nature of the high $T_{\rm B}$.

\section{Conclusions}\label{sec:conclusions}

In this paper, we report on the first 22\,GHz space VLBI observations of \m by {\it RadioAstron} in Feb 2014. The observations were scheduled with baseline lengths of up to $\sim11\,{\rm G}\lambda$ or at an equivalent interferometric angular resolution of $1/(11{\rm G}\lambda)\sim19\mu$as.
For the first time, interferometric fringes towards \m were detected on space baselines at 22\,GHz, up to $\sim3{\rm G}\lambda$, although this is much shorter than the planned maximum baseline length.
On longer baselines, adopting a threshold for the false fringe detection probability of $P_{e}\lesssim10^{-4}$, no significant fringes are found.

The new space VLBI observation yields the sharpest image of the source at 22\,GHz, which reveals a broad and edge-brightened jet and counterjet, whose geometry and brightness ratio are consistent with those from previous studies.
More importantly, the nucleus is well resolved into the north-south direction, revealing a brighter compact spot south of the core.
The minimum brightness temperature of the source was estimated from the visibility amplitude of the ground-to-space baseline, indicating high $T_{\rm B,min}\sim10^{12}\,$K. We have briefly discussed possible scenarios that could explain the exceptionally high brightness temperature in \m, including extreme Doppler boosting or intrinsically high $T_{\rm B,int}$ in the source. While combinations of various models do not contradict the observations, a unique interpretation is difficult to be chosen, due to the fact that \m is only modestly variable in radio and the jet is only mildly inclined towards the observer, compared to other extreme blazar and quasar jets that were also observed by {\it RadioAstron}. 
In this regard, analysis of further {\it RadioAstron} dataset of \m from the 2018 observation can be highly useful, in order to confirm the space fringe detection as well as time variability of the fine-scale structure in the core of the jet.

Finally, we remark that the main purpose of the 22\,GHz {\it RadioAstron} observation was to resolve the event-horizon scale structures in \m (e.g., \citealt{eht2019}). 
Similar to the {\it RadioAstron} observations of Sgr\,A* \citep{johnson21}, our study shows that observations at cm-wavelengths are fundamentally limited to low ground-to-space fringe detection rates, most likely due to the increased synchrotron opacity of the emitting plasma in the jet or accretion flow as well as scattering effects at long wavelengths. Therefore, it will be necessary for future space VLBI mission to observe at much higher radio frequencies (e.g., $\gtrsim$86\,GHz). Even though many technical challenges are expected, the significantly reduced opacity and much higher angular resolution will allow imaging of this unique source to reveal  unprecedented details of the lensed photon ring, infall and outflow of matter around active SMBHs (e.g., \citealt{andrianov21}).

\acknowledgments
We thank the anonymous referee for helpful comments which improved the paper and discussions, and Eduardo Ros for careful review of the manuscript and constructive suggestions.
JYK was supported for this research by the National Research Foundation of Korea (NRF) funded by the Korean government (Ministry of Science and ICT; grant no. 2022R1C1C1005255).
TS was funded by the Academy of Finland projects 274477 and 315721. 
EVK, YYK, APL were supported by the Russian Science Foundation project 20-62-46021.
S-SL was supported by the National Research Foundation of Korea (NRF) grant funded by the government of Korea (MIST) (2020R1A2C2009003). 
BWS is grateful for the support by the National Research Foundation of Korea funded by the Ministry of Science and ICT of Korea (NRF-2020K1A3A1A78114060).
The RadioAstron project was led by the Astro Space Center of the
Lebedev Physical Institute of the Russian Academy of Sciences and the
Lavochkin Scientific and Production Association under a contract with
the Russian Federal Space Agency, in collaboration with partner
organizations in Russia and other countries. The National Radio
Astronomy Observatory is a facility of the National Science Foundation
operated under cooperative agreement by Associated Universities,
Inc. The European VLBI Network is a joint facility of independent
European, African, Asian, and North American radio astronomy
institutes. This research is based on observations correlated at the
Bonn Correlator, jointly operated by the Max Planck Institute for
Radio Astronomy (MPIfR), and the Federal Agency for Cartography and
Geodesy (BKG).
The Australia Telescope Compact Array and Mopra telescope are part of the Australia Telescope National Facility (https://ror.org/05qajvd42) which is funded by the Australian Government for operation as a National Facility managed by CSIRO.

\vspace{5mm}
\facility{
{\it RadioAstron} Space Radio Telescope (Spektr-R), VLBA, EVN, JVLA, GBT, Effelsberg.
}


\software{
            AIPS \citep{greisen03},
            astropy \citep{astropy13}.
            Difmap \citep{shepherd94},
            eht-imaging \citep{chael16,chael18},
            ParselTongue \citep{kettenis06},
            PIMA\footnote{\url{http://astrogeo.org/pima/pima_user_guide.html}}.
          }

\clearpage

\newpage
\appendix
\section{Probability of false fringe detection with {\it RadioAstron}}\label{appendix:pfd}

It is well known that interferometric fringe detection is challenging in the low S/N regime, especially for space-VLBI, due to a large number of  uncertain parameters such as poorly determined position of the space antenna compared to the ground stations, which therefore introduce large residual fringe delay and rate errors. In such cases, the fringe finding algorithms can pick up a false signal from the broad fringe parameter space, even in the absence of the true signal. In general, the probability of the false fringe detection is theoretically understood for a given apparent S/N and is known in analytical forms (see, e.g., \citealt{tms}). However,
further systematic uncertainties, such as the antenna bandpass electronics and the exact definition of the S/N values in different algorithms, make it unavoidable to design and perform numerical simulations of the false fringes which greatly help overcome those limitations and obtain robust fringe detection statistics (e.g., \citealt{petrov11}).

In this work, we follow the approach of \cite{savolainen21} to better understand the reliability of the weak space fringe detection towards \m at $\sim3D_{\rm Earth}$
by AIPS FRING. %
We refer to their Appendix~A for the detailed setup of the false fringe simulations. In short, we shift the fringe search windows for the 3ED scan to regions of arbitrarily large parameters, where we are confident that no fringe solutions exist. Then, we compute the peak and its corresponding S/N values in the shifted window using the FFT module of AIPS FRING. Knowing that this peak is only a false detection, we can repeat similar random sampling of the peaks many times for different fringe window shifts. As a result, we can construct an empirical probability distribution of the false fringe detection versus specific S/N values as defined by AIPS FRING, as well as fully incorporating the data systematics.

\begin{figure}[t]
    \plotone{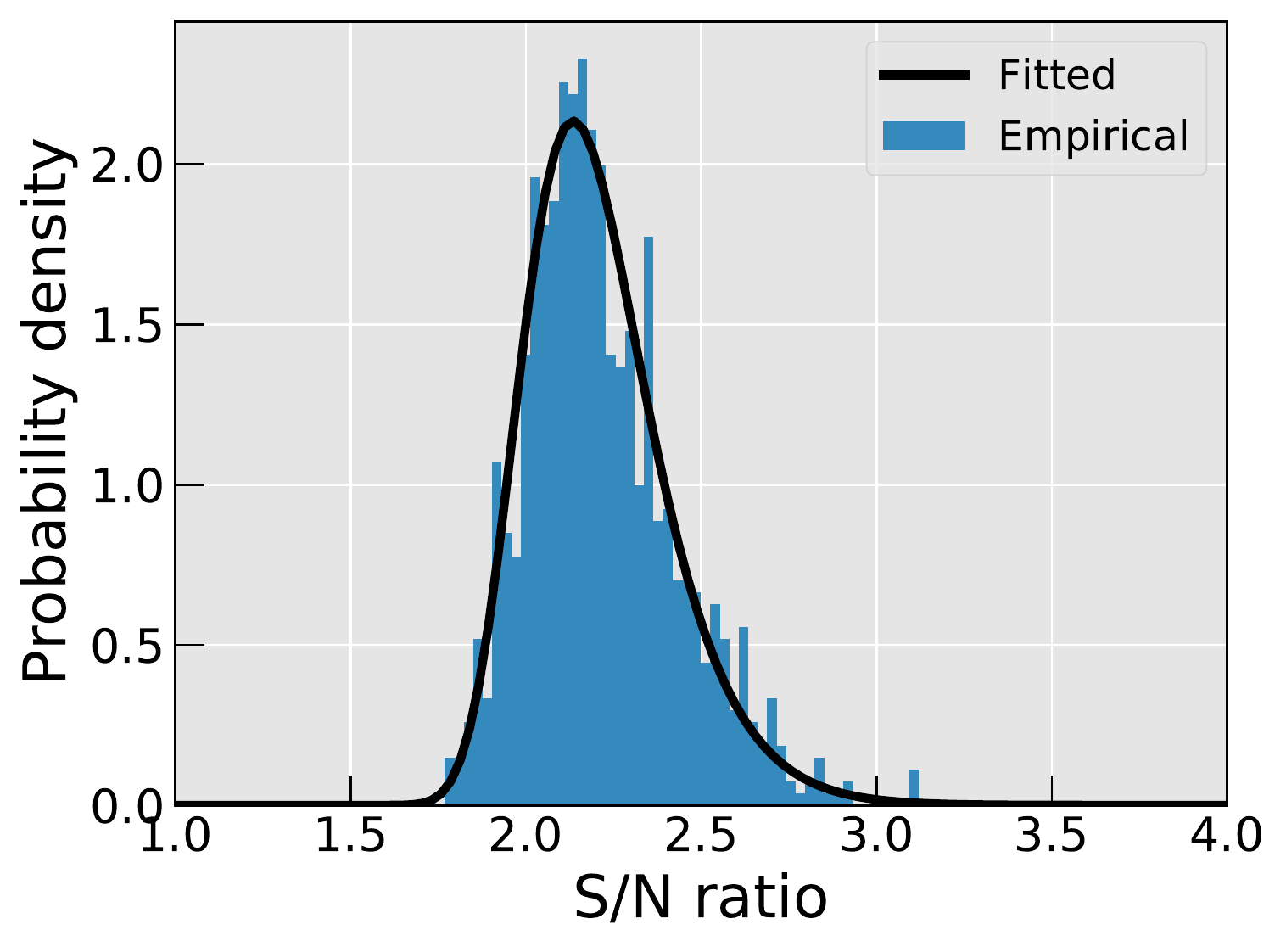}
    \caption{
    Note for the reviewer: we have removed title of this panel.
    Probability density distribution of the false fringe detection for baselines towards the SRT, as obtained by AIPS FRING, using the fringe search window of $\pm100$\,ns and $\,50$\,mHz.
    The blue bars show the empirical S/N distribution obtained from the Monte Carlo simulation with 1000 realizations of randomly large delay and rate offsets.
    The black solid line shows fit of Eq. \ref{eq:pfd} to the empirical distribution.
    }
    \label{fig:PFD}
\end{figure}

The result of the above Monte Carlo calculation is shown in Figure~\ref{fig:PFD}.
Then, a model for the probability density, $p(s)$, of finding a maximum fringe amplitude corresponding to a S/N value of $s$ within a specific fringe search window, when there is no true signal, is given \citep{tms,petrov11,savolainen21} by
\begin{equation}\label{eq:pfd}
    p(s)= \frac{n_{\rm eff}}{\sigma_{\rm eff}}f^{2}_{\rm SNR}s \times \exp\left(-\frac{(f_{\rm SNR}s)^{2}}{2}   \right)\times \left(1-\exp\left(-\frac{(f_{\rm SNR}s)^{2}}{2} \right) \right)^{n_{\rm eff}-1}~~,
\end{equation}
where $n_{\rm eff}$ is the effective number of grids in the Fourier space for the fringe search, $\sigma_{\rm eff}$ is the effective rms noise level of the fringe amplitude, and $f_{\rm SNR}$ is a correction factor accounting for the bias that the AIPS FRING task has when determining the $s$ value in the low S/N regime (see \citealt{desai98}).
We determine $n_{\rm eff}$, $\sigma_{\rm eff}$, and $f_{\rm SNR}$ by leaving them as free parameters and fitting Eq. \ref{eq:pfd} to the empirically derived PFD from the Monte Carlo simulation in Figure~\ref{fig:PFD}. Afterwards, we determine the probability of detecting a false fringe for a specific $s$ as reported by the AIPS FRING task, $P_{e}(s)$, by $P_{e}(s)=\int_{s}^{\infty}p(s')ds'$.
The results are tabulated in Table \ref{tab:pfd}.

\begin{figure}[t]
    \plotone{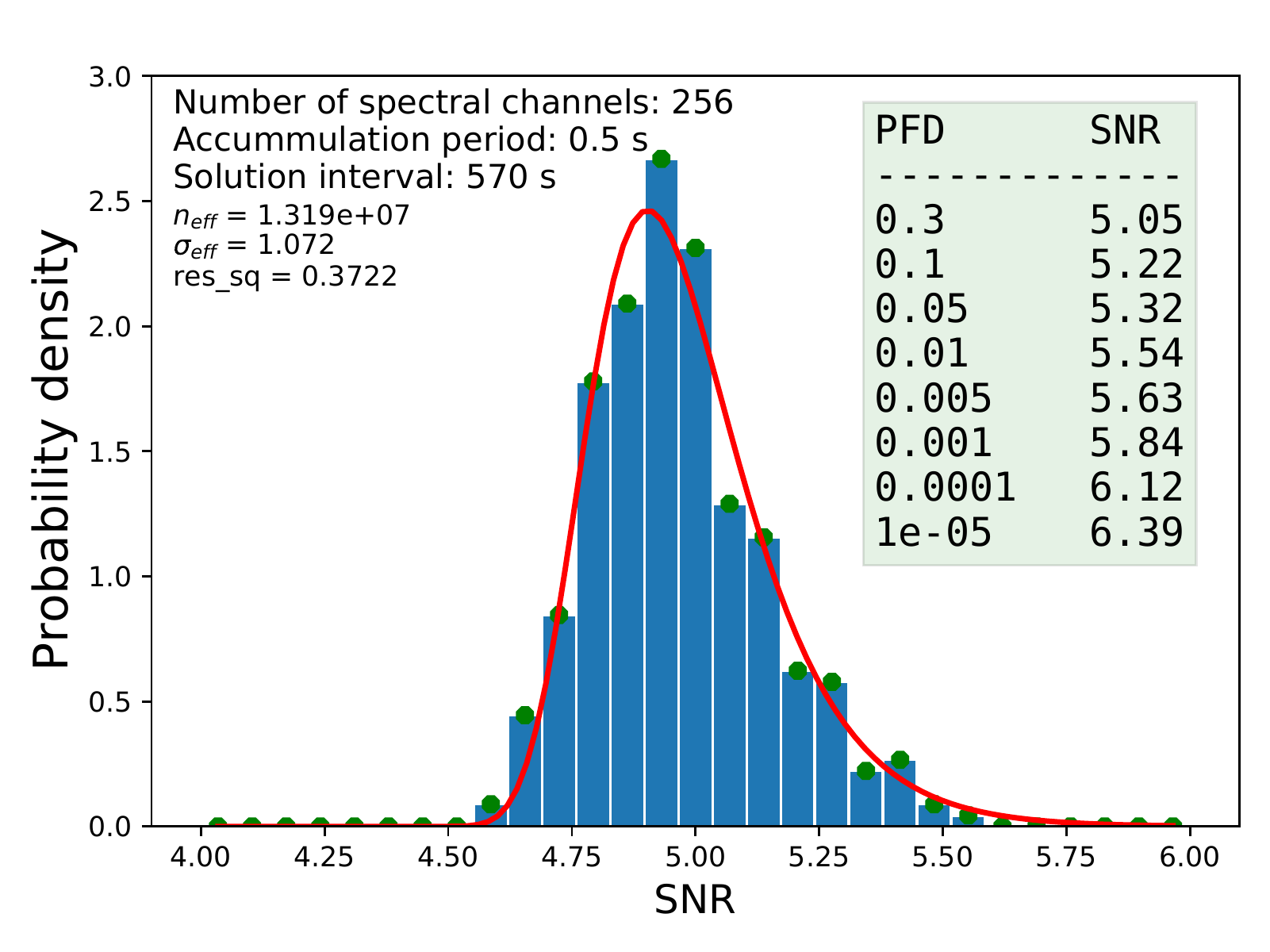}
    \caption{
    Note for the reviewer: the title has been removed.
    Probability density distribution of the false fringe detection for baselines towards the SRT for the LCP data, as obtained by PIMA.
    Inset shows parameters related to Eq. \ref{eq:pfd}.
    }
    \label{fig:PFD_pima}
\end{figure}

Essentially the same mathematical calculations are performed by PIMA \citep{petrov11,kovalev20} to obtain the fringe detection statistics and the results are shown in Figure~\ref{fig:PFD_pima} for a representative scan. Note that the exact definition of the S/N values is different in AIPS and PIMA, and thus the S/N values in Figures~\ref{fig:PFD} and \ref{fig:PFD_pima} are not identical.
For $P_{e}\sim10^{-4}$, we find an AIPS S/N threshold of $\sim3.3$ which  corresponds to the PIMA S/N threshold of $\sim6.1$.
The full PIMA fringe detection statistics for all the ground-space baselines, including the imaging and survey observations, are presented in Table \ref{table:flux_upper_limits}.

\begin{table}[]
    \centering
    \begin{tabular}{c c}
        S/N & $P_{e}$ \\
         \hline
        3.0 & $2.35\times10^{-3}$ \\
        3.1 & $9.58\times10^{-4}$ \\
        3.2 & $3.78\times10^{-4}$ \\
        3.3 & $1.45\times10^{-4}$ \\
        3.4 & $6.89\times10^{-5}$ \\
        3.5 & $2.49\times10^{-5}$ \\
        3.6 & $8.73\times10^{-6}$ \\
        3.7 & $3.89\times10^{-6}$ \\
        3.8 & $1.23\times10^{-6}$ \\
        3.9 & $4.14\times10^{-7}$ \\
         \hline
    \end{tabular}
    \caption{Probability of false fringe detection for a range of S/N values as defined and reported by AIPS FRING. The fringe search window has a size of $\pm100$\,ns and $\pm50$\,mHz.}
    \label{tab:pfd}
\end{table}

\section{Synthetic data imaging tests}\label{appendix:synthetic_data}

\begin{figure}[ht!]
    \plottwo{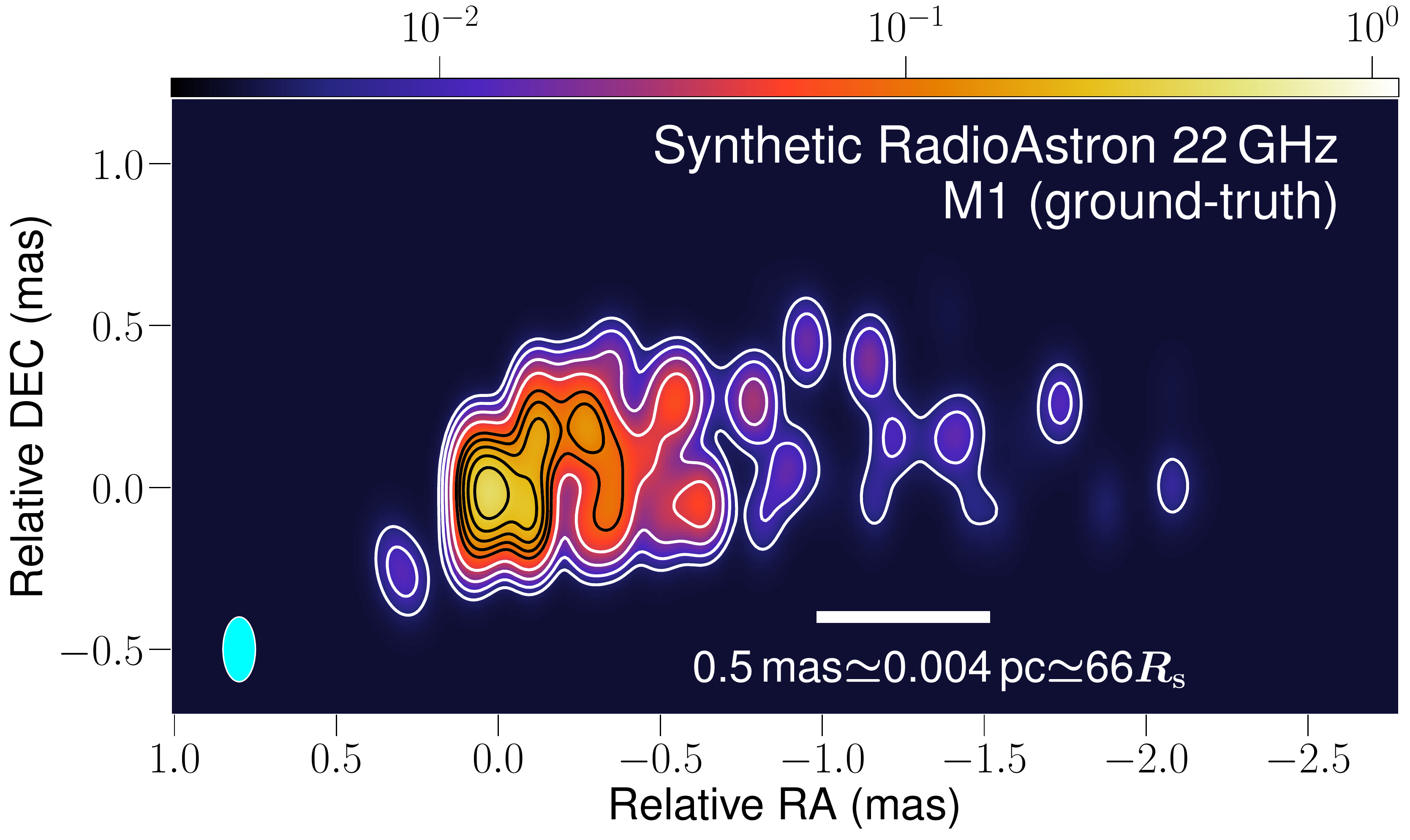}{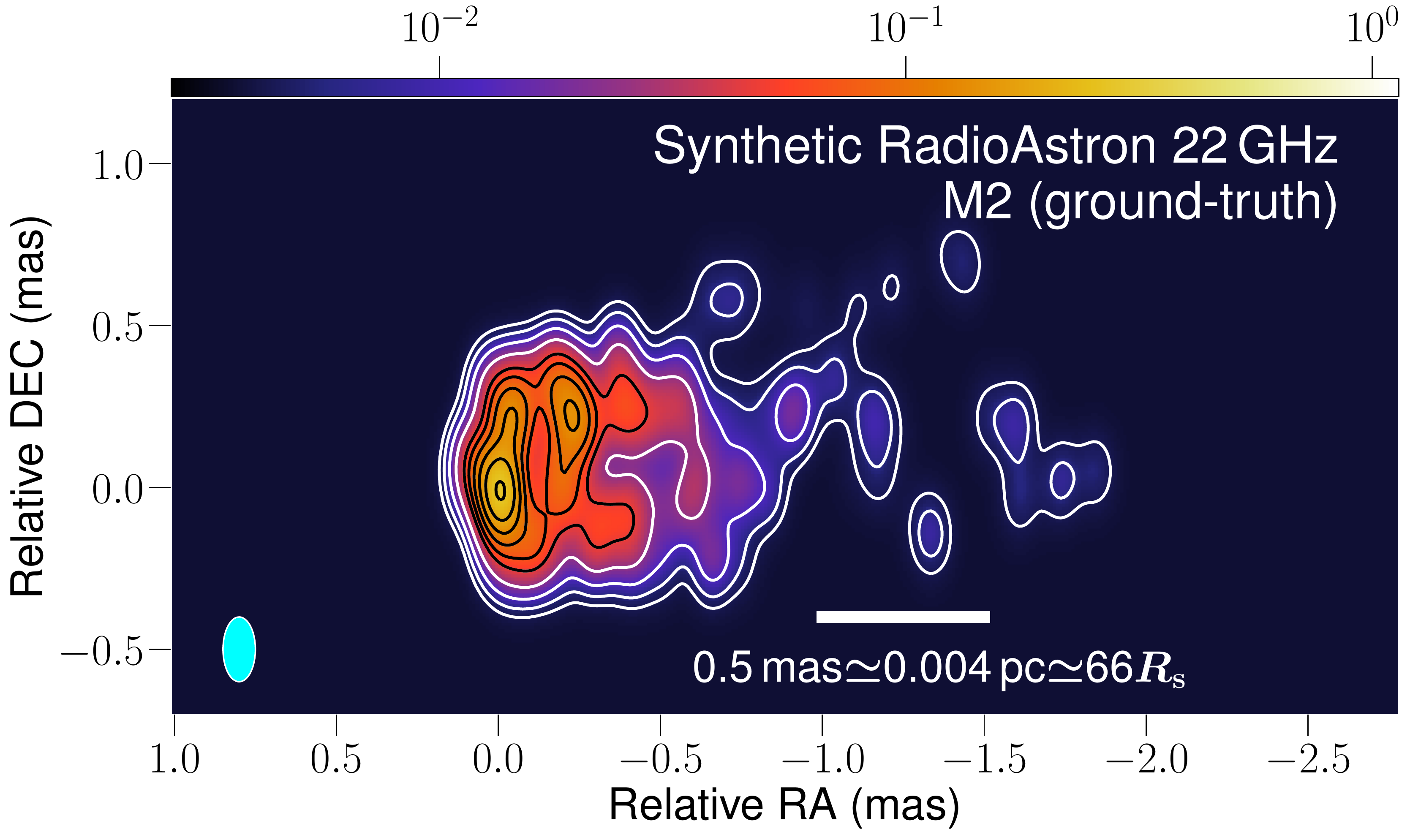}
    \plottwo{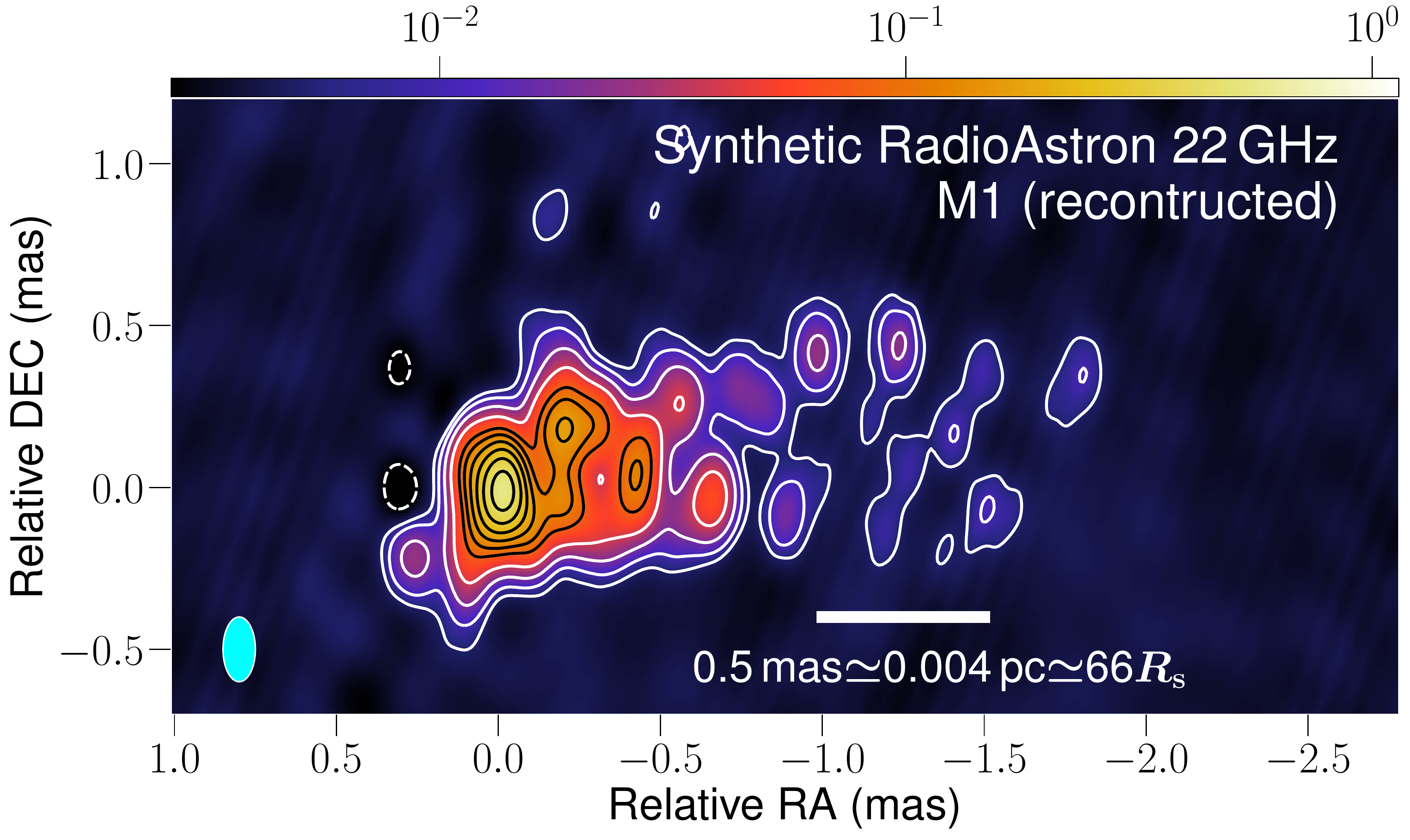}{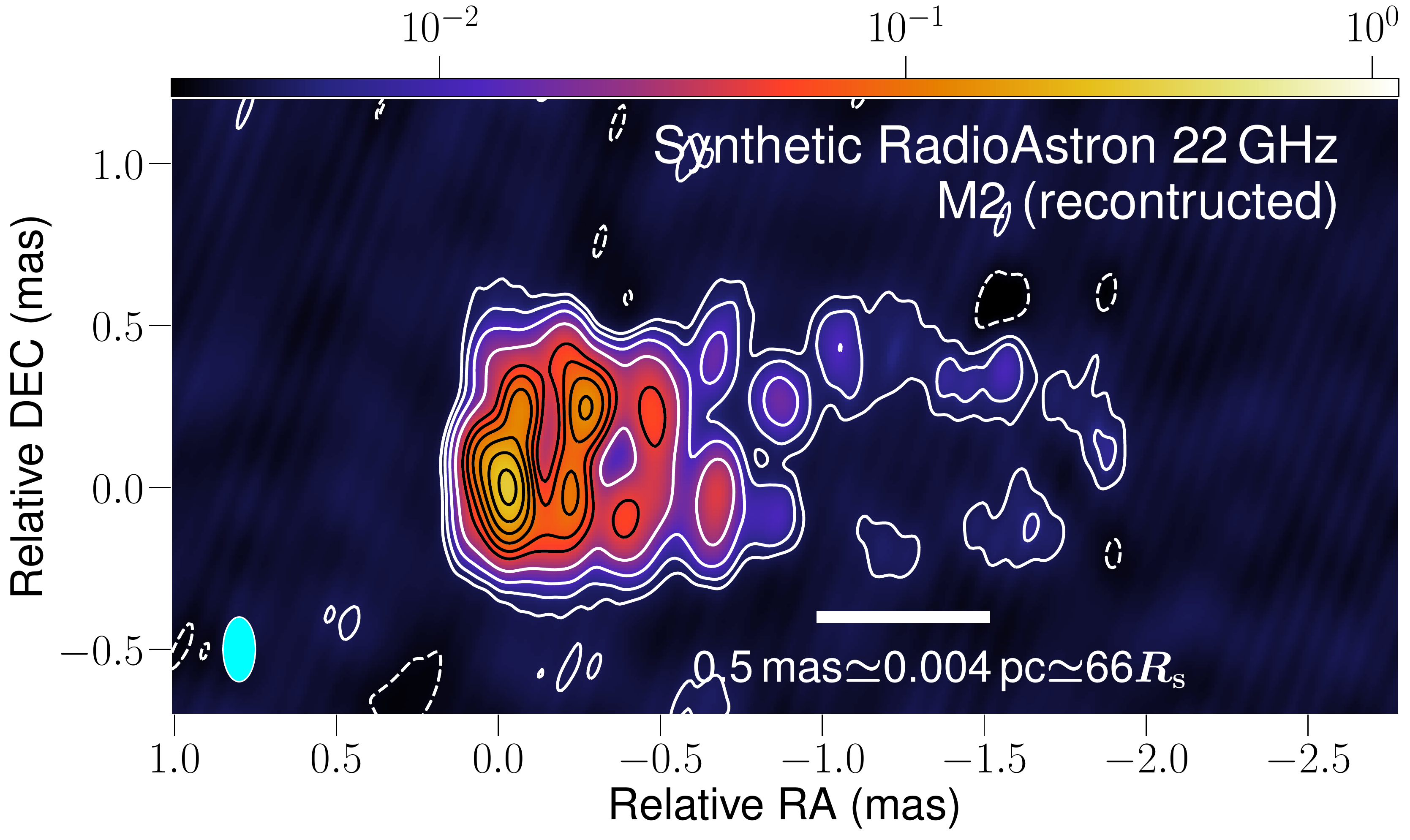}
    \plottwo{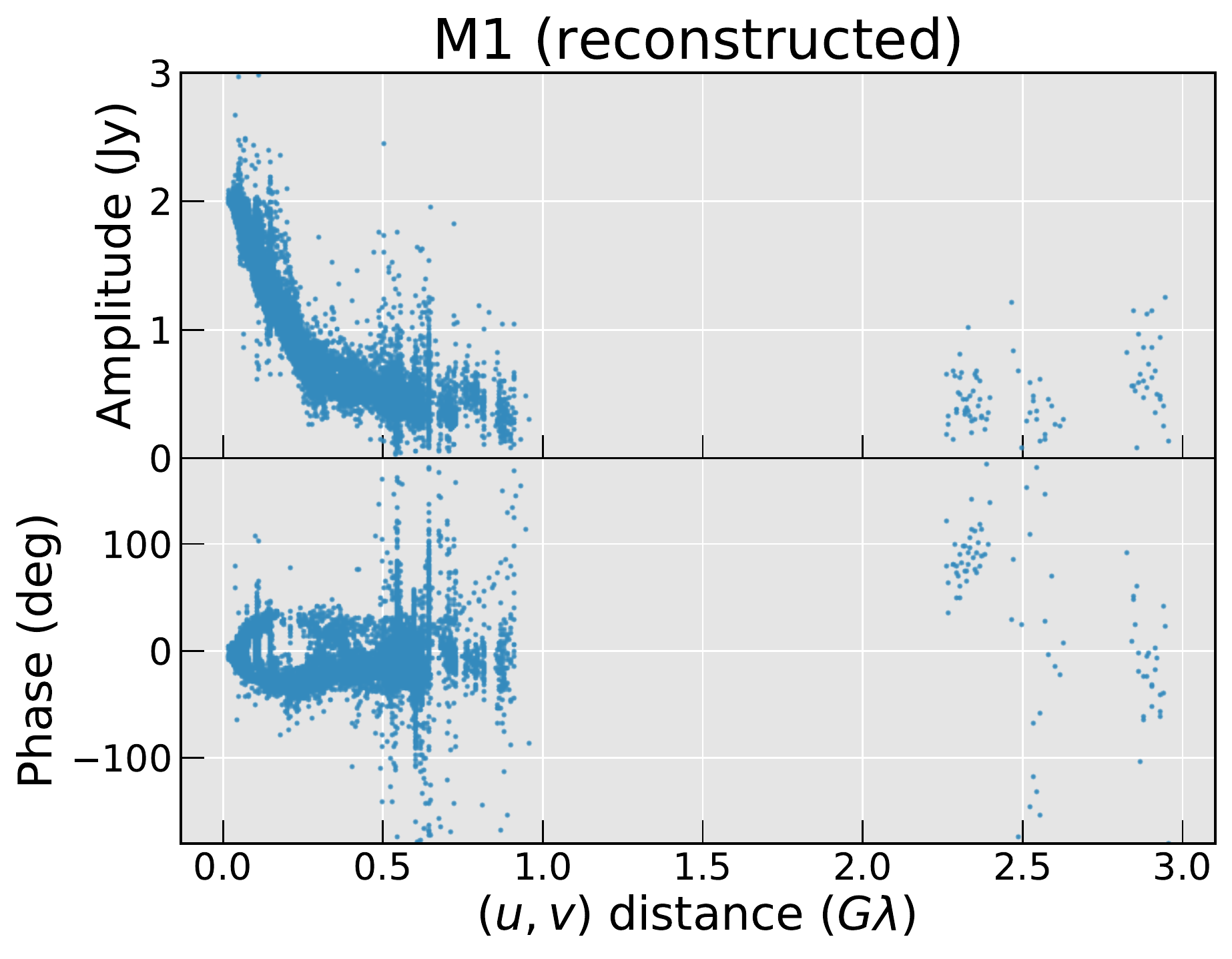}{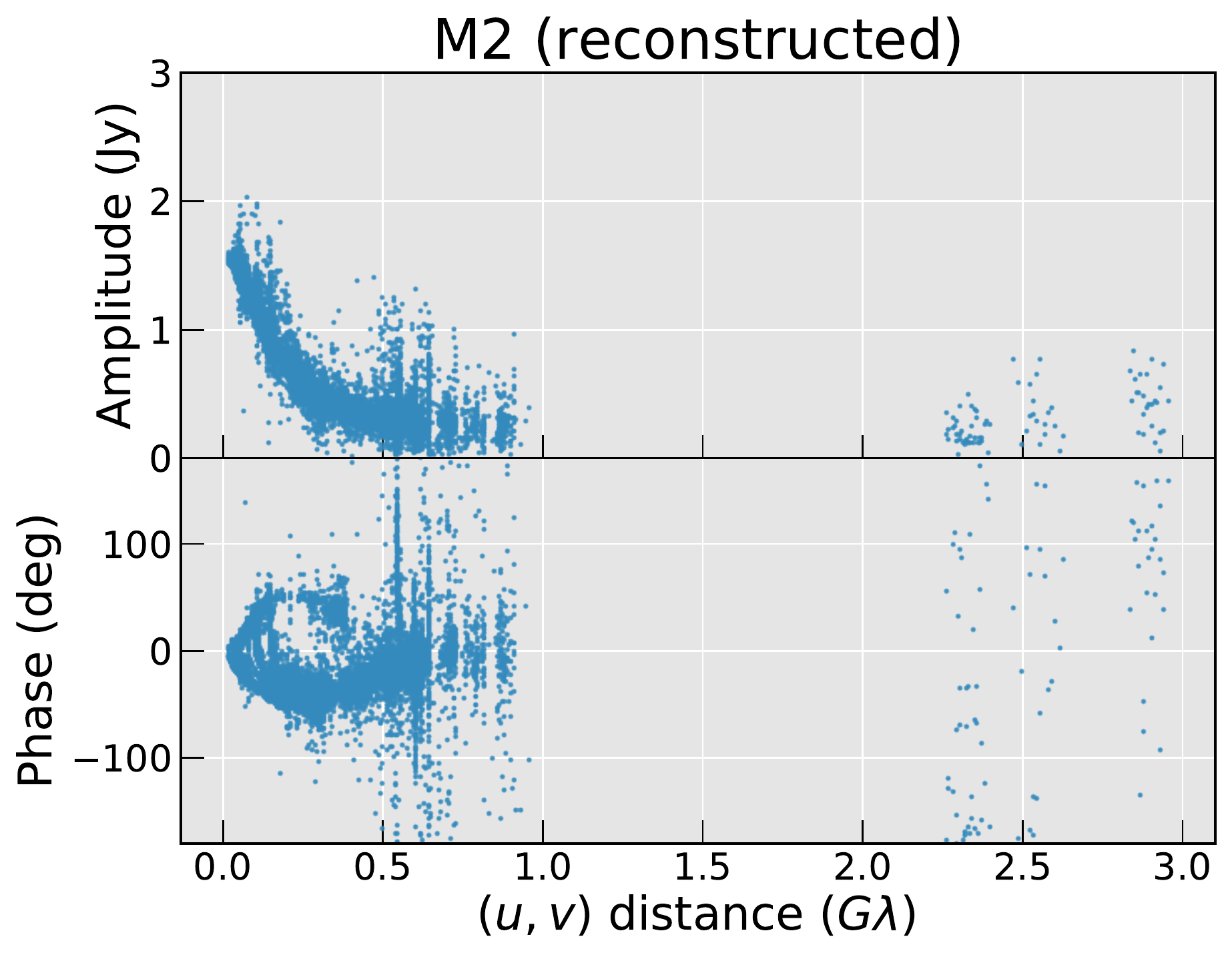}
    \caption{
    Summary of the synthetic data generation and imaging test.
    Top and middle rows show images of (left) M1 and (right) M2 models for the (top) ground-truth and (bottom) reconstructions.
    Here we adopt the same $0.2\times 0.1\,$mas beam as Figure~\ref{fig:final_images} bottom panel uses. Contours in the images for M1 (M2) start from 5\,mJy/beam (3\,mJy/beam) and increase by factor $\sqrt{2}$.
    The bottom row shows interferometric visibilities of (left) M1 and (right) M2 models after iterative CLEAN and phase self-calibrations. As in Figure~\ref{fig:amp_phs_uvdist}, the visibilities are averaged over 60\,s 
    and black/white contours are used for total intensity for better clarity. 
    }
    \label{fig:synthetic}
\end{figure}

Here we describe the overall procedure of our synthetic {\it RadioAstron} data generation and imaging tests, to evaluate the significance of the image reconstruction given the sparse $(u,v)$-coverage and limited S/N values of the ground-space baselines at 22\,GHz.
To begin with, we prepared two ground-truth models of \m, which consisted of CLEAN models from the imaging of the real dataset (\S\ref{sec:results}), but with slight modifications so that we do not reproduce exactly the same images in the test.
The first model, M1, was characterized by more compact, circular core region and fainter counterjet (Figure~\ref{fig:synthetic}, top left), with a total flux density of $\sim2.0$\,Jy. 
The second image, M2, contained no counterjet but broader nucleus in the N-S direction (Figure~\ref{fig:synthetic}, top right), with a total flux density of $\sim1.6$\,Jy.
Overall, these models were designed to test the imaging fidelity of the shape of the nuclear region and the presence of the counterjet.
Then we simulated mock observations of these models using the eht-imaging package \citep{chael16,chael18}, with exactly the same $(u,v)$-coverage as real observations. For the purpose of this test, we added to the synthetic observations only Gaussian random complex noises, using the signal-to-noise of each data point in the real observed data. The simulated visibilities were then imported into Difmap for imaging in a similar manner as real dataset was processed, with various CLEAN windows and other parameters such as uvweight, clean loop and gain, and phase self-calibrations (including the 1\,min solint limit for the spacecraft).
We note that the synthetic data imaging was also performed by multiple authors who were not informed about the underlying true intensity distribution.

The resulting images with slight super-resolution beam and reconstructed visibilities are all shown in Figure~\ref{fig:synthetic}. It can be seen by eye-comparison of the ground-truth models and reconstructed images that main features of the models such as the shape of the core, edge-brightening in the approaching jet, and the presence of counterjet are successfully recovered, although fainter details could vary.
On the other hand, we note that the reconstructed approaching jet shows discrepancies with the truth models, at distances $\gtrsim1.0\,$mas from the peak of the intensity. This is likely because of the super-resolving beam which resolves out too much the faint and extended structures. 
Therefore, we consider that the super-resolved image at the $2.0\times1.0$ beam is  reliable within $\sim1.0\,$mas from the peak of the intensity.

\section{Flux upper limits for the {\it RadioAstron} space baselines}\label{appendix:fluxes}

\startlongtable
\begin{deluxetable*}{cclllllrl}
\tablecaption{Upper limits on the flux densities of \m at 22\,GHz for space baselines to {\it RadioAstron}, in the order of increasing baseline length.
For each column (from left to right),
(1) $T_{\rm obs}$ is the starting time of the scans in year-month-date and hour-minute-second in UTC, 
(2) Experiment code for each observation (gs032 for the imaging experiment and others for the AGN survey observations),
(3) Name of the ground station,
(4) Solution interval used in the estimates of the S/N in seconds,
(5) and (6) observed and detection-threshold S/N values, respectively,
(7) baseline length in G$\lambda$,
(8) baseline position angle measured from north to east in degrees, and
(9) upper limits on the flux density in Jy.
Additional stations that participated in the AGN survey but not in the imaging observation (Table \ref{tab:station_list}) are DSS63 and VLA-N8, which are the Deep Space Station 63 antenna (Robledo 70\,m; geodetic code RO) and the phased JVLA (Y27; N8 for the phase center location). 
}
\startdata
$T_{\rm obs}$ & Expr. code & Station & $T_{\rm solint}$ & $S/N$ & $S/N_{\rm det}$ & $B$ & $PA$ & $F_{\rm upper}$ \\
 & & & (sec) & & & (G$\lambda$) & (deg) & (Jy) \\
\hline
2014-02-05 11:10:00 & gs032 & FD & 570 & 5.42 & 6.1 & 0.264 & 156.7 & 0.319 \\
2014-02-05 11:00:00 & gs032 & FD & 570 & 4.89 & 6.1 & 0.289 & $-$170.2 & 0.32 \\
2014-02-05 11:00:00 & gs032 & KP & 570 & 4.76 & 6.1 & 0.294 & 179.0 & 0.333 \\
2014-02-05 11:10:00 & gs032 & KP & 570 & 4.96 & 6.1 & 0.298 & 147.7 & 0.331 \\
2014-02-05 11:10:00 & gs032 & PT & 570 & 5.15 & 6.1 & 0.302 & 153.3 & 0.311 \\
2014-02-05 11:00:00 & gs032 & PT & 570 & 4.97 & 6.1 & 0.313 & $-$176.6 & 0.313 \\
2014-02-05 11:00:00 & gs032 & OV & 570 & 5.05 & 6.1 & 0.338 & 171.6 & 0.416 \\
2014-02-05 11:10:00 & gs032 & OV & 570 & 4.72 & 6.1 & 0.356 & 145.0 & 0.413 \\
2014-02-05 11:10:00 & gs032 & HN & 570 & 5.19 & 6.1 & 0.357 & $-$171.4 & 0.331 \\
2014-02-05 10:50:00 & gs032 & KP & 570 & 5.0 & 6.1 & 0.367 & $-$156.0 & 0.336 \\
2014-02-05 10:50:00 & gs032 & FD & 570 & 4.64 & 6.1 & 0.384 & $-$148.0 & 0.326 \\
2014-02-05 10:50:00 & gs032 & OV & 570 & 4.9 & 6.1 & 0.391 & $-$164.2 & 0.423 \\
2014-02-05 10:50:00 & gs032 & PT & 570 & 4.99 & 6.1 & 0.393 & $-$153.8 & 0.318 \\
2014-02-05 11:00:00 & gs032 & GB & 570 & 4.74 & 6.1 & 0.403 & $-$152.0 & 0.127 \\
2014-02-05 11:00:00 & gs032 & HN & 570 & 4.91 & 6.1 & 0.448 & $-$151.5 & 0.33 \\
2014-02-05 10:40:00 & gs032 & OV & 570 & 4.7 & 6.1 & 0.49 & $-$147.9 & 0.421 \\
2014-02-05 10:50:00 & gs032 & GB & 570 & 4.89 & 6.1 & 0.526 & $-$138.7 & 0.132 \\
2014-02-05 10:50:00 & gs032 & HN & 570 & 4.98 & 6.1 & 0.572 & $-$139.2 & 0.336 \\
2014-02-05 10:40:00 & gs032 & HN & 570 & 4.77 & 6.1 & 0.709 & $-$131.4 & 0.335 \\
2014-02-05 10:20:00 & gs032 & OV & 570 & 5.0 & 6.1 & 0.743 & $-$131.1 & 0.425 \\
2014-02-05 10:20:00 & gs032 & PT & 570 & 5.19 & 6.1 & 0.778 & $-$126.7 & 0.316 \\
2014-02-05 10:20:00 & gs032 & FD & 570 & 4.76 & 6.1 & 0.784 & $-$123.7 & 0.325 \\
2014-02-05 10:30:00 & gs032 & HN & 570 & 4.79 & 6.1 & 0.852 & $-$126.3 & 0.336 \\
2014-02-05 10:20:00 & gs032 & GB & 570 & 4.95 & 6.1 & 0.948 & $-$121.9 & 0.152 \\
2014-02-05 10:20:00 & gs032 & HN & 570 & 4.72 & 6.1 & 0.996 & $-$122.7 & 0.336 \\
2014-02-05 08:35:00 & gs032 & OV & 570 & 4.59 & 6.1 & 2.125 & $-$112.6 & 0.432 \\
2014-02-05 08:35:00 & gs032 & KP & 570 & 4.97 & 6.1 & 2.126 & $-$111.2 & 0.336 \\
2014-02-05 08:35:00 & gs032 & PT & 570 & 4.96 & 6.1 & 2.156 & $-$111.4 & 0.307 \\
2014-02-05 08:35:00 & gs032 & FD & 570 & 4.87 & 6.1 & 2.161 & $-$110.4 & 0.32 \\
2014-02-05 08:25:00 & gs032 & OV & 570 & 5.13 & 6.1 & 2.248 & $-$112.0 & 0.444 \\
2014-02-05 08:25:00 & gs032 & KP & 570 & 4.96 & 6.1 & 2.248 & $-$110.8 & 0.347 \\
2014-02-05 08:25:00 & gs032 & PT & 570 & 4.71 & 6.1 & 2.278 & $-$110.9 & 0.318 \\
2014-02-05 08:25:00 & gs032 & FD & 570 & 4.9 & 6.1 & 2.282 & $-$110.0 & 0.327 \\
2014-02-05 08:35:00 & gs032 & GB & 570 & 4.83 & 6.1 & 2.334 & $-$110.2 & 0.17 \\
2014-02-05 08:15:00 & gs032 & KP & 570 & 5.1 & 6.1 & 2.369 & $-$110.4 & 0.344 \\
2014-02-05 08:15:00 & gs032 & OV & 570 & 4.91 & 6.1 & 2.37 & $-$111.6 & 0.445 \\
2014-02-05 08:35:00 & gs032 & HN & 570 & 5.28 & 6.1 & 2.391 & $-$110.5 & 0.327 \\
2014-02-05 08:15:00 & gs032 & PT & 570 & 4.97 & 6.1 & 2.397 & $-$110.5 & 0.315 \\
2014-02-05 08:15:00 & gs032 & FD & 570 & 5.27 & 6.1 & 2.4 & $-$109.7 & 0.329 \\
2014-02-05 08:25:00 & gs032 & GB & 570 & 4.86 & 6.1 & 2.454 & $-$109.8 & 0.175 \\
2013-06-09 19:50:01 & raes03tf & EF & 569 & 5.45 & 6.7 & 2.487 & $-$96.4 & 0.159 \\
2014-02-05 08:25:00 & gs032 & HN & 570 & 4.94 & 6.1 & 2.512 & $-$110.1 & 0.337 \\
2013-06-09 20:00:02 & raes03tf & DSS63 & 560 & 5.56 & 6.7 & 2.53 & $-$102.4 & 0.134 \\
2013-06-09 19:50:03 & raes03tf & DSS63 & 559 & 5.65 & 6.7 & 2.549 & $-$97.9 & 0.143 \\
2014-02-05 08:15:00 & gs032 & GB & 570 & 5.09 & 6.1 & 2.571 & $-$109.4 & 0.174 \\
2014-02-05 08:15:00 & gs032 & HN & 570 & 5.03 & 6.1 & 2.63 & $-$109.7 & 0.334 \\
2014-02-05 08:35:05 & gs032 & YS & 570 & 4.97 & 6.1 & 2.703 & 71.1 & 0.246 \\
2014-02-05 08:25:06 & gs032 & YS & 570 & 5.1 & 6.1 & 2.832 & 71.6 & 0.251 \\
2014-02-05 08:15:01 & gs032 & EF & 570 & 4.94 & 6.1 & 2.946 & $-$109.6 & 0.162 \\
2014-02-05 08:15:07 & gs032 & YS & 570 & 4.96 & 6.1 & 2.959 & 71.9 & 0.258 \\
2014-02-05 06:50:00 & gs032 & KP & 570 & 5.01 & 6.1 & 3.347 & $-$108.2 & 0.346 \\
2014-02-05 06:50:00 & gs032 & FD & 570 & 4.93 & 6.1 & 3.364 & $-$107.7 & 0.325 \\
2014-02-05 06:50:00 & gs032 & OV & 570 & 4.82 & 6.1 & 3.364 & $-$108.9 & 0.457 \\
2014-02-05 06:50:00 & gs032 & PT & 570 & 4.81 & 6.1 & 3.371 & $-$108.2 & 0.306 \\
2014-02-05 06:40:00 & gs032 & KP & 570 & 4.84 & 6.1 & 3.458 & $-$108.0 & 0.356 \\
2014-02-05 06:40:00 & gs032 & FD & 570 & 4.79 & 6.1 & 3.473 & $-$107.6 & 0.333 \\
2014-02-05 06:40:00 & gs032 & OV & 570 & 4.83 & 6.1 & 3.478 & $-$108.7 & 0.472 \\
2014-02-05 06:40:00 & gs032 & PT & 570 & 5.05 & 6.1 & 3.481 & $-$108.1 & 0.314 \\
2014-02-05 06:50:00 & gs032 & GB & 570 & 5.01 & 6.1 & 3.512 & $-$107.5 & 0.102 \\
2014-02-05 06:30:00 & gs032 & KP & 570 & 5.11 & 6.1 & 3.568 & $-$107.8 & 0.358 \\
2014-02-05 06:50:00 & gs032 & HN & 570 & 4.85 & 6.1 & 3.572 & $-$107.7 & 0.329 \\
2014-02-05 06:30:00 & gs032 & FD & 570 & 5.07 & 6.1 & 3.581 & $-$107.4 & 0.332 \\
2014-02-05 06:30:00 & gs032 & OV & 570 & 4.75 & 6.1 & 3.59 & $-$108.5 & 0.475 \\
2014-02-05 06:30:00 & gs032 & PT & 570 & 5.03 & 6.1 & 3.591 & $-$107.9 & 0.312 \\
2014-02-05 06:40:00 & gs032 & GB & 570 & 5.19 & 6.1 & 3.617 & $-$107.4 & 0.105 \\
2014-02-05 06:40:00 & gs032 & HN & 570 & 4.68 & 6.1 & 3.677 & $-$107.5 & 0.336 \\
2014-02-05 06:30:00 & gs032 & GB & 570 & 4.86 & 6.1 & 3.721 & $-$107.2 & 0.106 \\
2014-02-05 06:30:00 & gs032 & HN & 570 & 4.91 & 6.1 & 3.781 & $-$107.4 & 0.338 \\
2014-02-05 06:50:06 & gs032 & YS & 570 & 5.42 & 6.1 & 3.948 & 74.0 & 0.257 \\
2014-02-05 06:50:01 & gs032 & SV & 570 & 4.87 & 6.1 & 3.96 & 71.9 & 0.385 \\
2014-02-05 06:50:01 & gs032 & TR & 570 & 5.31 & 6.1 & 3.979 & 72.7 & 0.573 \\
2014-02-05 06:40:07 & gs032 & YS & 570 & 5.04 & 6.1 & 4.055 & 74.2 & 0.262 \\
2014-02-05 06:40:01 & gs032 & SV & 570 & 4.97 & 6.1 & 4.075 & 72.1 & 0.399 \\
2014-02-05 06:40:01 & gs032 & TR & 570 & 4.81 & 6.1 & 4.092 & 72.9 & 0.553 \\
2014-02-05 06:30:06 & gs032 & YS & 570 & 4.89 & 6.1 & 4.16 & 74.3 & 0.264 \\
2014-02-05 06:30:01 & gs032 & EF & 570 & 4.62 & 6.1 & 4.178 & $-$106.7 & 0.146 \\
2014-02-05 06:30:01 & gs032 & SV & 570 & 4.98 & 6.1 & 4.188 & 72.4 & 0.397 \\
2014-02-05 06:30:01 & gs032 & TR & 570 & 4.69 & 6.1 & 4.202 & 73.1 & 0.505 \\
2014-02-05 05:05:00 & gs032 & GB & 570 & 4.86 & 6.1 & 4.579 & $-$106.4 & 0.138 \\
2014-02-05 05:05:00 & gs032 & HN & 570 & 4.96 & 6.1 & 4.632 & $-$106.4 & 0.329 \\
2014-02-05 04:55:00 & gs032 & GB & 570 & 4.98 & 6.1 & 4.677 & $-$106.3 & 0.165 \\
2014-02-05 04:55:00 & gs032 & HN & 570 & 4.98 & 6.1 & 4.73 & $-$106.4 & 0.337 \\
2014-02-05 04:45:00 & gs032 & GB & 570 & 4.76 & 6.1 & 4.776 & $-$106.2 & 0.174 \\
2014-02-05 04:45:00 & gs032 & HN & 570 & 4.87 & 6.1 & 4.826 & $-$106.3 & 0.338 \\
2014-02-05 05:05:06 & gs032 & YS & 570 & 4.76 & 6.1 & 4.993 & 75.2 & 0.272 \\
2014-02-05 05:05:01 & gs032 & TR & 570 & 5.01 & 6.1 & 5.081 & 74.4 & 0.348 \\
2014-02-05 04:55:06 & gs032 & YS & 570 & 4.8 & 6.1 & 5.084 & 75.3 & 0.291 \\
2014-02-05 05:05:01 & gs032 & SV & 570 & 5.21 & 6.1 & 5.091 & 73.8 & 0.387 \\
2014-02-05 04:55:01 & gs032 & EF & 570 & 5.2 & 6.1 & 5.133 & $-$105.4 & 0.128 \\
2014-02-05 04:45:05 & gs032 & YS & 570 & 5.22 & 6.1 & 5.175 & 75.4 & 0.29 \\
2014-02-05 04:55:01 & gs032 & TR & 570 & 4.72 & 6.1 & 5.178 & 74.5 & 0.35 \\
2014-02-05 05:05:01 & gs032 & ZC & 570 & 5.14 & 6.1 & 5.187 & 75.1 & 0.313 \\
2014-02-05 04:55:01 & gs032 & SV & 570 & 4.76 & 6.1 & 5.19 & 73.9 & 0.401 \\
2014-02-05 04:55:01 & gs032 & HH & 570 & 5.04 & 6.1 & 5.19 & $-$98.9 & 1.047 \\
2014-02-05 04:45:01 & gs032 & EF & 570 & 4.84 & 6.1 & 5.226 & $-$105.3 & 0.125 \\
2014-02-05 04:45:01 & gs032 & TR & 570 & 5.01 & 6.1 & 5.273 & 74.6 & 0.344 \\
2014-02-05 04:45:01 & gs032 & HH & 570 & 5.12 & 6.1 & 5.283 & $-$98.9 & 0.993 \\
2014-02-05 04:55:01 & gs032 & ZC & 570 & 4.98 & 6.1 & 5.286 & 75.2 & 0.31 \\
2014-02-05 04:45:01 & gs032 & SV & 570 & 4.74 & 6.1 & 5.288 & 74.1 & 0.401 \\
2014-02-05 04:45:01 & gs032 & ZC & 570 & 4.13 & 6.1 & 5.384 & 75.3 & 0.97 \\
2014-02-05 03:20:00 & gs032 & HN & 570 & 5.2 & 6.1 & 5.637 & $-$105.7 & 0.328 \\
2014-02-05 03:00:00 & gs032 & HN & 570 & 4.78 & 6.1 & 5.825 & $-$105.6 & 0.336 \\
2014-01-28 04:00:07 & raks01se & GB & 860 & 5.56 & 6.83 & 5.865 & 81.7 & 0.075 \\
2014-02-05 03:20:06 & gs032 & YS & 570 & 4.84 & 6.1 & 5.906 & 75.8 & 0.292 \\
2014-02-05 03:10:05 & gs032 & YS & 570 & 4.75 & 6.1 & 5.988 & 75.8 & 0.295 \\
2014-02-05 03:20:01 & gs032 & TR & 570 & 5.25 & 6.1 & 6.039 & 75.2 & 0.295 \\
2014-01-28 04:15:09 & raks01se & GB & 860 & 5.53 & 6.83 & 6.042 & 81.8 & 0.072 \\
2014-02-05 03:10:01 & gs032 & EF & 570 & 4.77 & 6.1 & 6.064 & $-$104.7 & 0.124 \\
2014-02-05 03:00:06 & gs032 & YS & 570 & 4.83 & 6.1 & 6.07 & 75.8 & 0.292 \\
2014-02-05 03:20:01 & gs032 & SV & 570 & 5.16 & 6.1 & 6.078 & 74.9 & 0.387 \\
2014-02-05 03:10:01 & gs032 & HH & 570 & 5.17 & 6.1 & 6.097 & $-$99.1 & 0.881 \\
2014-02-05 03:10:01 & gs032 & TR & 570 & 4.83 & 6.1 & 6.124 & 75.3 & 0.299 \\
2014-02-05 03:00:02 & gs032 & EF & 570 & 4.88 & 6.1 & 6.148 & $-$104.6 & 0.12 \\
2014-02-05 03:20:01 & gs032 & ZC & 570 & 5.0 & 6.1 & 6.159 & 76.0 & 0.256 \\
2014-02-05 03:10:01 & gs032 & SV & 570 & 5.06 & 6.1 & 6.166 & 74.9 & 0.399 \\
2014-02-05 03:00:01 & gs032 & HH & 570 & 4.94 & 6.1 & 6.176 & $-$99.1 & 0.851 \\
2014-02-05 03:00:01 & gs032 & TR & 570 & 4.91 & 6.1 & 6.208 & 75.3 & 0.299 \\
2014-01-28 04:30:08 & raks01se & GB & 861 & 5.8 & 6.83 & 6.217 & 81.9 & 0.108 \\
2014-02-05 03:10:01 & gs032 & ZC & 570 & 4.96 & 6.1 & 6.244 & 76.0 & 0.262 \\
2014-02-05 03:00:01 & gs032 & SV & 570 & 4.97 & 6.1 & 6.253 & 75.0 & 0.395 \\
2014-01-28 04:00:01 & raks01se & EF & 869 & 5.43 & 6.83 & 6.261 & 82.1 & 0.113 \\
2014-02-05 03:00:01 & gs032 & ZC & 570 & 4.88 & 6.1 & 6.329 & 76.1 & 0.261 \\
2014-01-28 04:00:01 & raks01se & HH & 868 & 5.6 & 6.83 & 6.37 & 87.4 & 0.816 \\
2014-01-28 04:15:02 & raks01se & EF & 838 & 5.77 & 6.83 & 6.454 & 82.1 & 0.238 \\
2014-01-28 04:15:02 & raks01se & HH & 839 & 5.6 & 6.83 & 6.566 & 87.3 & 0.854 \\
2014-02-05 01:35:06 & gs032 & YS & 570 & 5.25 & 6.1 & 6.744 & 76.0 & 0.233 \\
2014-02-05 01:25:05 & gs032 & YS & 570 & 4.76 & 6.1 & 6.822 & 76.0 & 0.245 \\
2014-02-05 01:25:02 & gs032 & HH & 570 & 4.91 & 6.1 & 6.889 & $-$99.3 & 0.735 \\
2014-02-05 01:35:01 & gs032 & TR & 570 & 4.78 & 6.1 & 6.898 & 75.7 & 0.251 \\
2014-02-05 01:15:05 & gs032 & YS & 570 & 4.76 & 6.1 & 6.9 & 76.1 & 0.237 \\
2014-02-05 01:25:01 & gs032 & EF & 570 & 4.9 & 6.1 & 6.914 & $-$104.3 & 0.108 \\
2014-02-05 01:15:01 & gs032 & HH & 570 & 5.38 & 6.1 & 6.96 & $-$99.4 & 0.714 \\
2014-02-05 01:35:01 & gs032 & SV & 570 & 4.92 & 6.1 & 6.962 & 75.5 & 0.334 \\
2014-02-05 01:25:01 & gs032 & TR & 570 & 4.77 & 6.1 & 6.977 & 75.7 & 0.256 \\
2014-02-05 01:15:01 & gs032 & EF & 570 & 4.96 & 6.1 & 6.992 & $-$104.3 & 0.105 \\
2014-02-05 01:35:01 & gs032 & ZC & 570 & 5.54 & 6.1 & 7.006 & 76.4 & 0.221 \\
2014-02-05 01:25:01 & gs032 & SV & 570 & 4.98 & 6.1 & 7.042 & 75.5 & 0.343 \\
2014-02-05 01:15:02 & gs032 & TR & 570 & 4.9 & 6.1 & 7.055 & 75.8 & 0.258 \\
2014-02-05 01:25:01 & gs032 & ZC & 570 & 4.78 & 6.1 & 7.082 & 76.5 & 0.229 \\
2014-02-05 01:15:01 & gs032 & SV & 570 & 4.99 & 6.1 & 7.122 & 75.6 & 0.345 \\
2014-02-05 01:15:01 & gs032 & ZC & 570 & 5.29 & 6.1 & 7.157 & 76.5 & 0.228 \\
2014-02-05 01:35:01 & gs032 & BD & 570 & 4.97 & 6.1 & 7.229 & $-$103.9 & 0.255 \\
2014-02-05 01:25:01 & gs032 & BD & 570 & 5.1 & 6.1 & 7.315 & $-$103.8 & 0.26 \\
2014-02-05 01:15:01 & gs032 & BD & 570 & 4.97 & 6.1 & 7.4 & $-$103.8 & 0.26 \\
2014-05-08 23:20:01 & raks01e4 & EF & 1199 & 5.71 & 6.92 & 8.043 & 92.9 & 0.126 \\
2014-05-08 23:20:00 & raks01e4 & KZ & 1200 & 5.64 & 6.92 & 8.1 & 92.5 & 0.521 \\
2014-05-08 23:00:03 & raks01e4 & EF & 1167 & 5.67 & 6.92 & 8.173 & 92.8 & 0.108 \\
2014-05-08 23:00:00 & raks01e4 & KZ & 1170 & 5.76 & 6.92 & 8.242 & 92.3 & 0.623 \\
2014-05-08 23:00:01 & raks01e4 & HH & 1140 & 6.29 & 6.92 & 8.397 & 96.5 & 0.996 \\
2014-02-04 20:10:00 & gs032 & MP & 570 & 4.89 & 6.1 & 9.741 & $-$99.0 & 0.366 \\
2014-02-04 20:10:00 & gs032 & AT & 570 & 4.68 & 6.1 & 9.747 & $-$99.0 & 0.109 \\
2014-02-04 20:00:00 & gs032 & MP & 570 & 4.66 & 6.1 & 9.802 & $-$99.0 & 0.338 \\
2014-02-04 20:00:00 & gs032 & AT & 570 & 4.97 & 6.1 & 9.807 & $-$99.0 & 0.101 \\
2014-02-04 18:25:00 & gs032 & MP & 570 & 4.84 & 6.1 & 10.338 & $-$99.0 & 0.347 \\
2014-02-04 18:25:00 & gs032 & AT & 570 & 4.79 & 6.1 & 10.343 & $-$99.1 & 0.11 \\
2014-02-04 18:15:00 & gs032 & MP & 570 & 5.29 & 6.1 & 10.392 & $-$99.0 & 0.329 \\
2014-02-04 18:15:00 & gs032 & AT & 570 & 5.12 & 6.1 & 10.396 & $-$99.1 & 0.1 \\
2014-02-04 16:46:05 & gs032 & AT & 570 & 4.07 & 6.1 & 10.854 & $-$99.2 & 0.219 \\
2014-02-04 16:40:00 & gs032 & MP & 570 & 4.71 & 6.1 & 10.882 & $-$99.2 & 0.338 \\
2014-02-04 16:30:00 & gs032 & MP & 570 & 4.85 & 6.1 & 10.932 & $-$99.2 & 0.314 \\
2015-02-14 02:07:34 & raks08qe & YS & 716 & 5.37 & 6.83 & 15.164 & 118.0 & 0.282 \\
2015-02-14 01:55:05 & raks08qe & YS & 715 & 5.35 & 6.83 & 15.184 & 117.9 & 0.275 \\
2015-02-14 02:07:32 & raks08qe & EF & 689 & 5.56 & 6.83 & 15.192 & 117.7 & 0.129 \\
2015-02-14 01:55:02 & raks08qe & EF & 718 & 5.47 & 6.83 & 15.215 & 117.5 & 0.126 \\
2015-02-14 02:07:32 & raks08qe & HH & 689 & 5.42 & 6.83 & 15.593 & 119.3 & 0.961 \\
2015-02-14 01:55:02 & raks08qe & HH & 718 & 5.53 & 6.83 & 15.61 & 119.2 & 0.957 \\
2013-02-03 13:30:03 & raes11b & VLA-N8 & 583 & 5.31 & 6.7 & 15.784 & 124.7 & 0.072 \\
2013-02-03 13:30:00 & raes11b & FD & 599 & 5.3 & 6.7 & 15.82 & 124.7 & 0.299 \\
2015-12-29 10:45:02 & raks12kj & EF & 898 & 5.62 & 6.83 & 20.244 & 128.4 & 0.142 \\
2015-12-29 10:30:02 & raks12kj & EF & 868 & 5.85 & 6.83 & 20.264 & 128.3 & 0.223 \\
2015-12-29 10:15:02 & raks12kj & EF & 868 & 5.39 & 6.83 & 20.283 & 128.3 & 0.128 \\
2016-01-07 09:15:01 & raks12lc & TR & 862 & 5.97 & 6.83 & 20.351 & 128.7 & 0.681 \\
2016-01-07 09:15:02 & raks12lc & EF & 568 & 5.8 & 6.7 & 20.364 & 128.8 & 0.293 \\
2015-02-12 03:45:02 & raks08pv & EF & 867 & 6.06 & 6.83 & 20.488 & 94.4 & 0.132 \\
2015-02-12 04:00:02 & raks08pv & EF & 839 & 5.56 & 6.83 & 20.497 & 94.5 & 0.11 \\
2015-02-12 03:45:01 & raks08pv & SV & 868 & 5.87 & 6.83 & 20.544 & 94.2 & 1.395 \\
2015-02-12 04:00:01 & raks08pv & SV & 868 & 5.48 & 6.83 & 20.545 & 94.3 & 0.372 \\
2015-02-12 03:45:02 & raks08pv & HH & 867 & 6.01 & 6.83 & 20.712 & 96.0 & 2.88 \\
2015-02-12 04:00:02 & raks08pv & HH & 838 & 5.37 & 6.83 & 20.724 & 96.0 & 0.805 \\
2013-12-29 08:15:07 & raks01ov & GB & 860 & 5.66 & 6.83 & 24.491 & 80.0 & 0.126 \\
2013-12-29 08:30:08 & raks01ov & GB & 861 & 5.83 & 6.83 & 24.491 & 80.0 & 0.319 \\
2013-12-29 08:00:07 & raks01ov & GB & 859 & 5.6 & 6.83 & 24.492 & 80.0 & 0.119 \\
2013-12-29 08:45:07 & raks01ov & GB & 891 & 5.84 & 6.83 & 24.492 & 80.0 & 0.194 \\
2013-12-29 08:45:01 & raks01ov & TR & 719 & 5.54 & 6.83 & 24.982 & 79.9 & 0.648 \\
2013-12-29 08:30:01 & raks01ov & TR & 862 & 5.55 & 6.83 & 24.988 & 80.0 & 0.544 \\
2013-12-29 08:15:01 & raks01ov & TR & 860 & 5.34 & 6.83 & 24.993 & 80.0 & 0.502 \\
2013-12-29 08:00:01 & raks01ov & TR & 858 & 5.54 & 6.83 & 24.997 & 80.0 & 0.476 \\
2014-02-01 04:00:06 & raks01sr & YS & 863 & 5.6 & 6.83 & 25.172 & 80.7 & 0.745 \\
2014-02-01 04:15:05 & raks01sr & YS & 864 & 6.14 & 6.83 & 25.193 & 80.7 & 0.279 \\
2014-02-01 04:00:01 & raks01sr & EF & 868 & 5.65 & 6.83 & 25.234 & 80.6 & 0.126 \\
2014-02-01 04:15:02 & raks01sr & EF & 839 & 5.67 & 6.83 & 25.251 & 80.6 & 0.132 
\enddata
\label{table:flux_upper_limits}
\end{deluxetable*}


\bibliography{8_m87_ra}{}
\bibliographystyle{aasjournal}



\end{document}